\documentclass[pra,superscriptaddress,showpacs,floatfix, longbibliography,twocolumn]{revtex4-2}

\usepackage[T1]{fontenc}
\usepackage[english]{babel}    
\usepackage{amsfonts}
\usepackage{amsmath}

\usepackage{graphicx,epsfig,epsf,color,hhline}% Include figure files
\usepackage{dcolumn}% Align table columns on decimal point
\usepackage{bm}% bold math
\usepackage{amsthm}
\usepackage{ bbold }
\usepackage{physics}
\usepackage{tensor,pbox}

\newcommand{\bea}{\begin{eqnarray}}
\newcommand{\ea}{\end{eqnarray}}
\newcommand{\eea}{\end{eqnarray}}

\newcommand{\bn}{{\bm n}}

\newcommand{\vc}{\ket{\text{vac}}_c}
\newcommand{\vaca}{\ket{\text{vac}}_a}

\newcommand{\vab}{{}_a\bra{\text{vac}}}

\newcommand{\Hx}{\hat{X}}
\newcommand{\Hy}{\hat{Y}}
\newcommand{\Hz}{\hat{Z}}

\newcommand{\Ta}{\hat{T}(\alpha)}
\newcommand{\Twa}{\hat{T}_\omega(\alpha)}
\newcommand{\Twad}{\dot{\hat{T}}_\omega(\alpha)}
\newcommand{\Swa}{\hat{S}_\omega(\alpha)}
\newcommand{\uom}{\hat{U}_\omega}
\newcommand{\tom}{\hat{T}_\omega}
\newcommand{\cH}{\mathcal{H}}
\newcommand{\cU}{\mathcal{U}}

\def\mbar{\bar{m}}

\newcommand{\pscal}[2]{\ensuremath{ \langle \, #1 \, \vert  \, #2 \, \rangle}} 
\newcommand{\dens}[2]{\ensuremath{ \vert \, #1 \, \rangle \langle \, #2 \, \vert}} 
\begin{document}
\title{Quantum parameter estimation with many-body fermionic systems and application to the Hall effect}

\author{Olivier Giraud}
\affiliation{Universit\'e Paris-Saclay, CNRS, LPTMS, 91405 Orsay, France}
\author{Mark-Oliver  Goerbig}
\affiliation{Universit\'e Paris-Saclay, CNRS, Laboratoire de Physique des Solides, 91405 Orsay, France.}
\author{Daniel Braun}
\affiliation{Institut f\"ur theoretische Physik, Universit\"at T\"{u}bingen, 72076 T\"ubingen, Germany}

\begin{abstract}
 We calculate the quantum Fisher information for a generic many-body
 fermionic system in a pure state depending on a parameter. We discuss
 the situations where the parameter is imprinted in the basis states,
 in the state coefficients, or both. In the case where the parameter
 dependence of coefficients results from a Hamiltonian evolution, we
 derive a particularly simple expression for the quantum Fisher
 information. We apply our findings to the quantum Hall effect, and
 evaluate the quantum Fisher information associated with the optimal
 measurement of the magnetic 
 field for a system in the ground state of the effective
 Hamiltonian. 
 The occupation of electron states with high momentum
 enforced by the Pauli principle leads to a super-Heisenberg scaling
 of the sensitivity with a power law that depends on the geometry of
 the sensor. 
 \end{abstract}

%%%%%%%%%%%%%%%%%%%%%%%%%%%%%%%%%%%%%%%%%%%%%%%%%%%%%%

\date{March 17, 2023}

\maketitle

%*************************************************************************
%*************************************************************************
\section{Introduction}
%*************************************************************************
%*************************************************************************
The Hall effect offers  a precise and economic way of
measuring magnetic fields with small, integrated sensors.  Typical
commercially available Hall sensors 
based on silicon have sensitivities of about
100\,nT/Hz$^{1/2}$ \cite{Roelver2013}.
Graphene-based ones
are projected to achieve sensitivities normalized to the width $w$ ($B_\text{min}w$) of 4\,pT$\cdot\text{mm}/\sqrt{\text{Hz}}$ at room temperature \cite{xu_batch-fabricated_2013}.  The 
quantum Hall effect, reached at very strong magnetic fields and low
temperatures, has also become a cornerstone of metrology, allowing a
measurement of the von Klitzing constant $R_K=h/e^2$ to 10 digits precision \cite{Codata2018}.  \\

In the present work we do not investigate the precision with  which one can access $R_K$, but assess the
ultimate sensitivity of magnetic field sensors based on the quantum
Hall effect.  This ultimate sensitivity is only bound by quantum
noise and thermal noise of the sensor, and should be attainable once
all the technical noises, such as electrical noise in the
amplifiers and wires, vibrations, fluctuating charges in the materials
etc.~have been removed. A powerful formalism for calculating this
ultimate sensitivity is provided by  
the quantum Cram\'er-Rao bound (QCRB)
\cite{helstrom_quantum_1969,Holevo1982,Braunstein1994}, expressed in
terms of the quantum 
Fisher information (QFI), which leads to an important 
ultimate 
benchmark of the
sensitivity. \\

Motivated by the Hall effect application, we first investigate here more generally
quantum parameter estimation with a system consisting of a large number of
indistinguishable fermions (typically electrons). Such a system is
most concisely described by fermionic quantum field theory, which we
will briefly review in the following for setting up the notations used. 
We will consider quantum states written in a basis of many-particle states.
These basis states are obtained by ``creating'' fermions in single-particle states, chosen here as eigenstates of some single-particle Hamiltonian.  We will consider three different ways by which a parameter dependence can be imprinted on such a state: via a parameter-dependent evolution Hamiltonian, via parameter-dependent fermionic many-particle basis states, or via a Bogoliubov transformation. Note that many other possibilities exist to imprint a parameter on a state, see \cite{braun2018quantum}. In all three cases mentioned above, a parameter-independent initial state will be transformed into a parameter-dependent state via a parameter-dependent unitary transformation. Let us now discuss these three ways in turn.

{\em i}.) Time evolution generated by a Hamiltonian that depends on a parameter is the standard situation in most single-particle or single-mode
applications of quantum metrology, where one
typically considers a fixed (parameter-independent) basis and initial
states, and time-dependent
and parameter-dependent amplitudes 
for those.
Both dependencies of the amplitudes arise from propagation with the Hamiltonian.

{\em ii}.) The parameter dependence may be imprinted in the many-particle basis states themselves.  Indeed, as these are anti-symmetrized linear combinations
of single-particle energy eigenstates, a change of the single-particle
Hamiltonian can modify its eigenbasis, so that even without
time evolution the state of the system can contain information about
the value of the parameter.  For a specific example, consider a single
particle in a harmonic trap. Assume the particle is in the
ground state of the  oscillator, and the parameter we are interested
in is the frequency of the trap.  Through the oscillator length the
ground-state wave function clearly depends on the frequency.
Increasing the frequency squeezes the ground-state wave function in
position space. Hence, even without time-evolution, one can measure,
at least in principle, the frequency of the harmonic oscillator (see
\cite{braun2011ultimate} for details). In quantum optics, the
quantum fluctuations 
of the vacuum state (i.e.~without any photons present) have indeed been
measured directly \cite{riek_direct_2015}, and it 
is clear that they depend on the frequency considered.

An underlying physical assumption of this reasoning is that the system
is always in the actual
parameter-dependent ground state (or any other state specified through
a given number of 
excitations of a single-particle Hamiltonian or linear combinations
thereof) when the parameter is changed. Of course, if the system
remained in the ground state corresponding to some fixed value
$\omega_0$ of the parameter $\omega$, no dependence on
$\omega$ arises, and $\omega$ 
cannot be measured without time evolution. Hence, a
relaxation mechanism is required when reducing the ground-state
energy, and energy has to be pumped into the system when the energy
increases as function of $\omega$. A more complete description of the
system should take into account the relaxation mechanism and/or pumping mechanism.
The time scale on which the corresponding
information is imprinted on the state might
compete with the time scales of the evolution of the state driven by
the Hamiltonian.  However, according to the formalism of the
quantum Cram\'er-Rao bound, only infinitesimal changes of the parameter
need to be considered for determining the best sensitivity with which
the parameter can be estimated, and hence only an infinitesimal amount
of relaxation or pumping suffices to justify the model.  We will
therefore assume that the 
system indeed tracks the parameter-dependent many-particle states
instantaneously over infinitesimal changes of the parameter without
specifying the physical mechanism that allows it to do so. 

{\em iii.}) {In the more general situation encountered in
quantum field theory} the number of particles need not
be conserved, which creates an additional freedom for encoding parameters
compared to single-particle quantum mechanics.  Indeed, the most
general linear transformations of the creation and annihilation
operators that preserve their fermionic anti-commutation relations are
Bogoliubov transformations.  We will therefore consider Bogoliubov
transformations as a third way of coding parameters in a state.
{In the most general situation, Bogoliubov transformations allow
 to mix excitations with creation of holes, which opens the way to a
 new kind of quantum parameter estimation not possible with
 single-particle basis change. We will first discuss this general case
 and derive very general expressions for the QFI. We will then
 consider the special case where the particle number is conserved, that is, when Bogoliubov transformations mix creation
operators with creation operators only and annihilation operators with
annihilation operators only. This corresponds to changing the single-particle basis states. This case will be relevant to application of our results to the quantum Hall effect.}

Bogoliubov transformations for quantum
parameter-estimation have been considered
before in bosonic field theories \cite{PhysRevD.89.065028}. Analytical
results were obtained for the estimation of small parameters in terms of Bogoliubov
coefficients for single-mode and two-mode Gaussian channels.  The QFI for specific two-mode bosonic Gaussian states was also
found in \cite{PhysRevA.93.052330}.  In
\cite{vsafranek2015quantum} an exact expression for the QFI of an arbitrary two-mode bosonic Gaussian state was
obtained.  Carollo and co-workers
calculated the symmetric logarithmic
derivative of general Gaussian fermionic states
\cite{carollo_symmetric_2018}. In \cite{PhysRevA.89.032326} a proper definition
of entanglement in fermionic systems and its connection to the
sensitivity of quantum metrology schemes based on them was investigated.

Here, we investigate
quantum-parameter estimation for arbitrary pure states of
indistinguishable fermions, and include all three ways of encoding a
parameter described above. Performing a time evolution,
a basis change or a Bogoliubov transformation amounts to applying a
unitary operator to the initial quantum state. In Section \ref{sec:QFI} we calculate the
QFI in the case where an initial
parameter-independent state is subjected to a parameter-dependent
unitary transformation. We then derive a chain rule for the QFI
in the case of two successive unitary
transformations, which allows us to identify the contribution from
each of them as well as their mutual influence.  In Section \ref{specCases} we calculate the QFI for
Bogoliubov transformations (whose formalism is reviewed in
Section \ref{formalism}). Section \ref{sec:qhe} is dedicated to applying this formalism
to the quantum Hall effect.

%*************************************************************************
%*************************************************************************
\section{Fermionic quantum field theories and Bogoliubov transformations}
\label{formalism}
%*************************************************************************
%*************************************************************************

%*************************************************************************
\subsection{Fermionic basis states}
%*************************************************************************
The most general pure state of indistinguishable fermions in $M$
single-particle modes can be written in  the form
\begin{equation}
  \label{eq:psifer}
  \ket{\psi}=\sum_{\bm n}\psi_{\bm n}\ket{\bm n}_c\,,
\end{equation}
where the sum runs over all $M$-tuples $\bn=(n_0,n_1,\ldots,n_{M-1})$ with $n_k\in\{0,1\}$, and 
\begin{equation}
  \label{eq:nc}
\ket{\bm n}_c=\prod_{k=0}^{M-1}(c_k^\dagger)^{n_k}\vc 
\end{equation}
is the state of $n_k$ particles in mode $k$ for
$k=0,\ldots,M-1$. Here, the states $\ket{\bm n}_c$ are the
$N$-particle states $\ket{n_0}_c\otimes\ket{n_1}_c\otimes\cdots\otimes\ket{n_{M-1}}_c$, or equivalently $\ket{n_0,n_1,\ldots,n_{M-1}}_c$, with $N=\sum_k n_k$, and $\ket{n_k}_c$ are eigenstates of a
single-particle Hamiltonian, with
$k=0,1,2,\ldots$.  The operator $c_k^\dagger$ creates a fermion in 
%the single-particle state $\ket{k}_c$
mode $k$ out of the vacuum $\vc$. The vacuum
state $\vc$ of particles $c$ is defined as the state that satisfies 
$c_k\vc=0$ $\forall k=0,\ldots,M-1$, i.e.~it is a state that contains no
particles of type $c$. 

The set of all  $2^M$ states $\ket{\bm n}_c$, with $\bm n$ running over all
$M$-tuples of 0 and 1, forms a basis of Fock space, and  
$\psi_{\bm n}$ in \eqref{eq:psifer} are the (complex) coefficients of
$\ket{\psi}$ in that basis. We will consider $\ket{\psi}$ to
depend on a parameter $\omega$ that we want to estimate. In the most general situation the
parameter dependence can arise both from the $\psi_{\bm n}$ and from the
basis states $\ket{\bm n}_c$. 
Note that for energies much
smaller than the rest masses of the 
fermions, superpositions containing a different number of particles
are forbidden by the particle-number superselection rule.  States of
the form \eqref{eq:psifer} are nevertheless considered for example in BCS
theory of superconductivity \cite{PhysRev.108.1175}, where particle number conservation is
enforced only on average (and to a very good precision, for a large number of
particles). Of course, by an appropriate choice of the $\psi_{\bm
  n}$, one can restrict $\ket{\psi}$ to a
state with a   
fixed number of particles.   

%*************************************************************************
\subsection{Bogoliubov transformations}  \label{bogtra}
%*************************************************************************
We consider the situation where the $c_k$ arise from a
parameter-dependent Bogoliubov transformation from parameter-independent creation and annihilation operators $a_k^\dagger$ and $a_k$. Bogoliubov transformations are the most general linear
transformations that preserve canonical
anticommutation 
relations. They take the general form
\begin{eqnarray}
  \label{eq:Bog}
  c_i^\dagger=a_j^\dagger U_{ji}+a_j V_{ji}\nonumber\\
  c_i=a_j U_{ji}^*+a_j^\dagger V_{ji}^*
\end{eqnarray}
 (with
Einstein summation convention), where $U_{ji}$, $V_{ji}$ are parameter-dependent complex numbers.  
The preservation of the anticommutation relations $\{c_i^\dagger,c_j\}=\{a_i^\dagger,a_j\}=\delta_{ij}$ implies the condition
$U^\dagger U+V^\dagger V=\text{Id}_M$, while $\{c_i,c_j\}=\{a_i,a_j\}=0$
gives $U^tV+V^tU=0$, where $V^t$ denotes the transpose of $V$, and $\text{
  Id}_M$ the $M\times M$-dimensional identity matrix.
When arranged as a matrix $W$ with
\begin{equation}
  \label{eq:W}
  W=\begin{bmatrix}
U & V^*\\
V& U^*
\end{bmatrix}\,,
\end{equation}
the two above conditions on $U$ and $V$ can be equivalently expressed
as $W^\dagger W=\text{Id}_{2M}$,
so that $W$ is unitary. Following \cite{takayanagi_utilizing_2008} we
introduce the compact 
vector notation $\alpha=(a^\dagger, \,a)\equiv (a^\dagger_0,\ldots
a^\dagger_{M-1},a_0,\ldots a_{M-1})$, and correspondingly $\gamma=(c^\dagger,
\,c)\equiv (c^\dagger_0,\ldots  
c^\dagger_{M-1},c_0,\ldots c_{M-1})$.  The Bogoliubov transformation
\eqref{eq:Bog} can then 
be written simply as $\gamma=\alpha W$. 

Let $S$ be the $2M\times 2M$ matrix defined from $W$ by the relation
\begin{equation}
  \label{eq:Sdef}
W=\exp(i S \Xi)  
\end{equation}
 with
\begin{equation}
  \label{eq:X}
  \Xi=
  \begin{bmatrix}
    0&\text{ Id}_M\\
\text{ Id}_M & 0
  \end{bmatrix}\,.
\end{equation}
Because of \eqref{eq:Sdef} and the definition \eqref{eq:W}, the matrix
$S$ has the block form 
\begin{equation}
  \label{eq:S}
S=  \begin{bmatrix}
    S^{(2)}& S^{(1)}\\
-S^{(1)*}& - S^{(2)*}
  \end{bmatrix}\,,
\end{equation}
where $S^{(1)}$ and $S^{(2)}$ are (in general complex) $M\times M$
matrices, with $S^{(1)}=S^{(1)\dagger}$ Hermitian and
$S^{(2)}=-S^{(2)t}$ antisymmetric. The matrix $S$ is not Hermitian in general, but it satisfies $S^t=-S$, and hence $S^\dagger=-S^*$. 
We define the operators
\begin{align}
  \label{eq:Shat}
\hat{S}(\alpha)&=\frac{1}{2}\alpha S \alpha^t\,, \\
\hat{T}(\alpha)&=\exponential(i \hat{S}(\alpha))\,.
 \label{eq:T}
\end{align}
Since $\alpha^t= \Xi\alpha ^\dagger$ (where the $^\dagger$ conjugates the annihilation and
creation operators and transforms the row vector into a column
vector), $\hat{S}(\alpha)$ can be written in the alternative form
$\hat{S}(\alpha)=\frac12\alpha S \Xi \alpha^\dagger$. 
The operator $ \hat{T}(\alpha)$ satisfies the identity
\begin{equation}
  \label{eq:Bog2}
\hat{T}(\alpha)\alpha\hat{T}(\alpha)^\dagger=\alpha W=\gamma\,.
\end{equation}

%*************************************************************************
\subsection{Relation between bases}
%*************************************************************************
To the vacuum state $\vc$ for particles of type $c$ corresponds a vacuum
state $\vaca$ for particles of type $a$. It is defined
by $a_k\vaca=0$ $\forall k=0,\ldots,M-1$. In general, the two vacua are  
different, $\vc\ne \vaca$, as is obvious from the fact that whenever
$V_{ji}\neq 0$ in Eq.~\eqref{eq:Bog}, the operator $c_i$ creates a
particle of type $a$.  The two vacua are related via 
\begin{equation}
  \label{eq:vcva}
  \vc=\hat{T}(\alpha)\vaca\,,
\end{equation}
as can be readily seen by noting that Eqs.~\eqref{eq:Bog2} and \eqref{eq:vcva} imply
\begin{equation}
  \label{eq:proof1}
  c_k\vc=\hat{T}(\alpha)a_k \hat{T}^\dagger(\alpha)\hat{T}(\alpha)\vaca=
  \hat{T}(\alpha)a_k\vaca=0
\end{equation}
(see e.g.~\cite{takayanagi_utilizing_2008}). Only in the case where $V=0$
(i.e.~the Bogoliubov transformation does not mix creation operators
with annihilation operators) does one have, up to possibly a phase,
$\vc=\vaca$. 
Equation \eqref{eq:vcva} generalizes to an arbitrary state: one has
for a Fock state
\begin{eqnarray}
  \label{eq:proofpsi}
\ket{\bm n}_c&=&\prod_{k=0}^{M-1}(c_k^\dagger)^{n_k}\vc\nonumber\\
&=&\prod_{k=0}^{M-1}\left(\Ta(a_k^\dagger)^{n_k}\Ta^\dagger\right)
    \Ta\vaca\nonumber\\
&=&\Ta\prod_{k=0}^{M-1}(a_k^\dagger)^{n_k}\vaca\nonumber\\
&=&\Ta\ket{\bm n}_a\,,
\end{eqnarray}
and by linearity for an arbitrary pure state
\begin{eqnarray}
  \label{eq:proofpsi2}
\ket{\psi}= \sum_{\bm n}\psi_{\bm n}\ket{\bm n}_c&=&\Ta\sum_{\bm
                                                  n}\psi_{\bm
                                                  n}\ket{\bm
                                                  n}_a\,.
\end{eqnarray}

%*************************************************************************
\subsection{One-particle overlaps}
%*************************************************************************
We now calculate the overlap between
one-particle states in terms of the Bogoliubov parameters. Let $R$ be
the $M\times M$ matrix defined as 
\begin{equation}
\label{eq:R}
R_{kl}={}_a\braket{k}{l}_c\,,
\end{equation}
where $\ket{k}_c$ is the state with one particle in mode $k$, i.e.~the state $\ket{\bm n}_c$ with $n_i=\delta_{ik}, 0\leq i\leq M-1$. Using the expression of $\gamma_i$ given by Eq.~\eqref{eq:Bog2}, we have for $i,j\in\{1,\ldots,2M\}$
\begin{equation}
\label{eq:sand}
  \vab\alpha_j\,\gamma_i\vc=\vab\alpha_j\,\alpha_k\vc W_{ki}
\end{equation}
(again with implicit summation).
Since by definition $a_j\vaca=\ket{0}$ and $c_i\vc=\ket{0}$, the
left-hand side of \eqref{eq:sand} has the block structure 
\begin{equation}
  \label{eq:0R00}
  \begin{bmatrix}
    0&0\\R&0\,
  \end{bmatrix}\,.
\end{equation}
On the right-hand side of \eqref{eq:sand} the term $\vab\alpha_j\,\alpha_k\vc$ has the block structure
\begin{equation}
  \label{eq:00MN}
  \begin{bmatrix}
    0&0\\P&Q\,
  \end{bmatrix}.
\end{equation}
Using the block structure  \eqref{eq:W} of $W$, Eq.~\eqref{eq:sand} readily gives
\begin{align}
\label{pupv1}
PU+QV&=R\\
PV^*+QU^*&=0\,.
\label{pupv2}
\end{align}
Matrices $P$ and $Q$ can be calculated by using the following
canonical decomposition for the operators $\Ta$
\cite{takayanagi_utilizing_2008}:  
\begin{equation}
\label{Tcanon}
\Ta=|U^\dagger|^{1/2}e^{\Hz}e^{\Hy}e^{\Hx}\,,
\end{equation}
where
\begin{eqnarray}
\Hx&=&\frac12\sum_{i,j}X_{ij}a_i a_j,\quad X=U^{*-1}V,\\
\Hy&=&\frac12\sum_{i,j}Y_{ij}a_i^\dagger a_j,\quad e^{-Y}=U^\dagger,\\
\Hz&=&\frac12\sum_{i,j}Z_{ij}a_i^\dagger a_j^\dagger, \quad Z=V^{*}U^{*-1}\,,
\label{defZ}
\end{eqnarray}
and $|.|$ denotes the determinant (recall that in general $U$ is not a
unitary matrix). While operators $\Hx$ and $\Hy$ contain annihilation
operators, $\Hz$ only contains creation operators. Thus
$\vc=\Ta\vaca=|U^\dagger|^{1/2}e^{\Hz}\vaca$ and the overlap between
vacua reads 
\begin{equation}
\label{overlapvac}
\vab\text{vac}\rangle_c=|U^\dagger|^{1/2}\,.
\end{equation}
Matrix $P$ is readily obtained as
\begin{equation}
\label{matP}
P_{kj}=\vab a_k\,a_j^\dagger\vc=|U^\dagger|^{1/2}\delta_{kj}\,.
\end{equation}
Using the identity \cite{onishi_generator_1966}
\begin{equation}
\label{onishi}
\left[a_k,e^{\Hz}\right]=\sum_lZ_{kl}a_l^\dagger e^{\Hz}
\end{equation}
(which can be shown by induction), we get
\begin{eqnarray}
\label{overlapaaZ}
\vab a_n a_k e^{\Hz}\vaca&=&\vab a_n \left[a_k,e^{\Hz}\right]\vaca\nonumber\\
&=&\sum_l\vab a_n a_l^\dagger e^{\Hz}\vaca Z_{kl}\nonumber\\
&=&\vab e^{\Hz}\vaca Z_{kn}\nonumber\\
&=&-Z_{nk}
\end{eqnarray}
(the relation $U^tV+V^tU=0$ implies $Z^t=-Z$). Therefore,
\begin{equation}
\label{matQ}
Q_{nk}=\vab a_n a_k \vc=-Z_{nk}|U^\dagger|^{1/2}\,.
\end{equation}
When we replace $P$ and $Q$ by their above expression in Eq.~\eqref{pupv2}, we get $V^*-ZU^*=0$, which is in fact a direct consequence of
the definition of $Z$. Doing the same in Eq.~\eqref{pupv1} we get $|U^\dagger|^{1/2}(U-ZV)=R$.
Using \eqref{defZ}, this gives 
\begin{equation}
\label{schur1}
R=|U^\dagger|\left(U-V^*U^{*-1}V\right).
\end{equation}
We recognize the Schur complement of the block $U^*$ in matrix
$W$, which appears in the expression for the inverse of the block matrix $W$. Since $W$ is unitary, Eq.~\eqref{schur1} reduces to
\begin{equation}
\label{schur2}
R=|U^\dagger|\,U^{\dagger-1}\,.
\end{equation}
If the Bogoliubov transformation is such that $V=0$, then the relation
$U^\dagger U+V^\dagger V=\text{Id}_M$ implies that $U$ is
unitary. From Eq.~\eqref{schur2} we then have $U=R$, so that for this particular
Bogoliubov transformation  $U$ is simply the matrix of
one-particle overlaps. In other words, the Bogoliubov transformation
between two single-particle bases can be obtained by taking $U=R$ and $V=0$.  
This is the situation we encounter in section \ref{urv0}
 below.

%*************************************************************************
%*************************************************************************
\section{Quantum Cram\'er-Rao bound and quantum Fisher information
  in fermionic quantum field theories} 
\label{sec:QFI}
%*************************************************************************
%*************************************************************************

%*************************************************************************
\subsection{Quantum Cram\'er-Rao bound}
%*************************************************************************
Let $\ket{\psi_\omega}$ be a quantum state which depends on a parameter
$\omega$. More generally, consider a density matrix $\rho(\omega)$ that
describes a parameter-dependent mixed state. One would like to know 
how precisely one can estimate $\omega$ based on the measurement of
some observables.  This will depend in general,
of course, on a lot of things, starting with the measurements chosen,
the precision of the measurement devices used, the noisiness of the
environment, the number of 
measurements, the statistics of the data obtained, and how the data are
analyzed.  However, with the quantum Cram\'er-Rao bound (QCRB)
\cite{helstrom_quantum_1969,Holevo1982,Braunstein1994}, a very powerful theoretical tool is
available that allows one to calculate the smallest possible
uncertainty of any unbiased estimate of $\omega$, no matter what
positive-operator-valued measure (POVM) measurements are performed,
and what estimator functions are used to analyze the data, as long as
they are unbiased estimator functions based on the measurement results
alone. Suppose we want to estimate a parameter $\omega$ by measuring
$M_e$ times a quantity $X$ (e.g.~a POVM) whose statistics of outcomes
$P(X=x|\omega)$ depends on $\omega$. An estimator
$\hat{\omega}(x_1,\ldots,x_{M_e})$ is any function that maps the 
$M_e$ 
measurement results $x_1,\ldots,x_{M_e}$ to an estimate of the parameter $\omega$. It is called unbiased if 
the average of $\hat{\omega}$ for the probability distribution
$P(X=x|\omega)$ is $\langle \hat{\omega}\rangle=\omega$ locally. With
such an estimate at hand, measurement of $X$ allows one to access
$\omega$. However, since the measurement results fluctuate in general
due to the quantum nature of the state, so does the estimator. Its
smallest possible variance 
gives 
the optimal sensitivity with which one can 
estimate $\omega$ by
measuring $X$. The QCRB provides a lower bound for the variance of
$\hat{\omega}$. It reads 
\begin{equation}
  \label{eq:qcrb}
  \text{var}(\hat{\omega})\ge\frac{1}{M_e}\frac{1}{I(\rho(\omega),\omega)}\,, 
\end{equation}
where $I$ is the quantum Fisher information, given by
\begin{equation}
  \label{eq:qfi}
  I(\rho(\omega),\omega)=\tr(\rho(\omega)L^2(\omega))\,
\end{equation}
and the symmetric logarithmic derivative operator $L(\omega)$ is a linear
operator defined through
\begin{equation}
  \label{eq:Lom}
  \partial_\omega
  \rho(\omega)=\frac{1}{2}(L(\omega)\rho(\omega)+\rho(\omega)L(\omega))\,. 
\end{equation}
The bound can be saturated in the limit of $M_e\to\infty$.
In \cite{Braunstein94} it was shown that the QFI can be interpreted
geometrically through the Bures distance between two states
$\rho(\omega)$ and $\rho(\omega+\delta\omega)$ that differ infinitesimally
in the parameter. 
This gives an appealing physical interpretation to
the QCRB: the ultimate sensitivity with which a parameter coded in a
quantum state can be estimated is all the more large as the state depends strongly
on the parameter. 

%*************************************************************************
\subsection{General expressions for the quantum Fisher information}
%*************************************************************************
The QFI for systems with infinite-dimensional Hilbert space is in
general difficult to calculate, as it typically requires the
diagonalization of the density matrix in order to determine the
logarithmic derivative or the calculation of the
Bures distance. However, if the state is given already in diagonalized
form, closed expressions for the QFI can be found.  The simplest case
in this category is that of a pure state
$\ket{\psi_\omega}$. Its QFI can be shown to be \cite{Braunstein96}
\begin{equation}
  \label{eq:I}
  I(\ket{\psi_\omega},\omega)=4\left(
\braket*{\dot\psi_\omega}{\dot\psi_\omega}+\braket*{\dot\psi_\omega}{\psi_\omega}^2
\right)\,,
\end{equation}
where $\ket*{\dot\psi_\omega}\equiv
\partial_\omega\ket{\psi_\omega}$ (see Eq.~(26) in
\cite{Paris09}). Note that in the whole paper dots denote derivatives with respect to the parameter $\omega$.
For a mixed state $\rho(\omega)$ given in its eigenbasis, 
$\rho(\omega)=\sum_r p_r(\omega)
\dens{\psi^{(r)}_\omega}{\psi^{(r)}_\omega}$, where the 
$\ket*{\psi^{(r)}_\omega}$ form an orthonormal basis, one has
\begin{equation}\label{QFI_diag}
I(\rho(\omega),\omega) =\sum_r \frac{(\partial_\omega p_r)^2}{p_r} +2
\displaystyle\sum_{\substack{n,m }}\frac{(p_n-p_m)^2}{p_n+p_m} \left|
  \pscal{\psi^{(n)}_\omega}{\dot\psi^{(m)}_\omega} \right|^2  \;, 
\end{equation} 
where the sums are over all terms with
non-vanishing denominators.

{The form \eqref{eq:I} can be equivalently expressed as
  \begin{equation}
    I(\ket{\psi_\omega},\omega)=4\bra*{\dot\psi_\omega}\left(\mathbb{1}-\ketbra*{\psi_\omega}\right)\ket*{\dot\psi_\omega}.
  \end{equation}
Under that form, the QFI can be directly related to the Fubini-Study metric. More generally, the QFI has a simple geometric interpretation: it is related to the Bures distance between two infinitesimally close states \cite{Hubner1992} via the identity \cite{JulienThesis2017}
  \begin{equation}\label{QFIBures}
I(\rho(\omega),\omega) =4\lim_{\delta\omega\to 0}\frac{d_B\left(\rho(\omega),\rho(\omega+\delta\omega)\right)^2}{\delta\omega^2},
\end{equation}
with
\begin{equation}\label{QFIBures}
d_B(\rho,\rho')=\left(2-2\tr\sqrt{\rho^{\frac12}\rho'\rho^{\frac12}}\right)^{\frac12}.
\end{equation}
}
%*************************************************************************
\subsection{QFI for a unitary transformation} \label{single}
%*************************************************************************

\subsubsection{General pure state}
The most general pure states of a system described within quantum field theory are of the form \eqref{eq:psifer}. Both the basis states $\ket{\bm n}_c$ and the amplitudes $\psi_{\bm  n}(\omega)$ can depend on the parameter $\omega$, so that we have to deal with states of the form
\begin{equation}
  \label{eq:psigeneral}
  \ket{\psi_\omega}=\sum_{\bm n} \psi_{\bm n}(\omega)\ket{\bm
    n}_\omega\,. 
\end{equation}
The reason for this is that the basis states $\ket{\bm n}_\omega$ are
constructed as antisymmetrized linear combinations of
single-particle eigenstates that can depend on the parameter through
the single-particle Hamiltonian. For example, in the case of the
quantum Hall effect that we will consider in detail in section
\ref{sec:qhe}, the single-particle energy eigenstates correspond to
Landau levels that depend on the magnetic field (or equivalently the
cyclotron frequency $\omega$), i.e.~they are
energy eigenstates of an harmonic oscillator with frequency $\omega$,
leading e.g.~in position representation to wavefunctions with a
typical length scale given by the frequency-dependent oscillator
length.  In addition, the propagation of a 
superposition of eigenstates leads to parameter-dependent phases of
the amplitudes. 
As we will show, the form \eqref{eq:psigeneral} can be obtained from a parameter-independent state by means of two consecutive unitary operators. 
We first consider the case where a single unitary operator is applied. Of course, one could always combine these two unitaries
into a single one, but for some applications the decomposition into
two unitaries is natural, as will be illustrated in Sec.~\ref{twounitexample} below.

\subsubsection{A single unitary}
Suppose the unitary transformation is of the form
$\hat{T}_\omega=\exponential(i\hat{S}_\omega)$, with $\hat{S}_\omega$
Hermitian, acting on some parameter-independent reference state
$\ket{\psi_{\omega_0}}$, so
that the state is of the form
$\ket{\psi_\omega}=\hat{T}_\omega\ket{\psi_{\omega_0}}$. This
situation arises for example by a time evolution driven
by a 
Hamiltonian $\hat{H}_\omega$ that depends on the parameter $\omega$ (in which case $\hat{S}_\omega=-\hat{H}_\omega t$ is also proportional to time $t$). 
A simple calculation shows that the QFI \eqref{eq:I} can be rewritten as  
\begin{equation}
\label{qfiH}
I(\ket{\psi_\omega},\omega)=4\,\text{var}(\mathcal H, \ket{\psi_{\omega_0}})\,,\qquad\textrm{with }{\mathcal H}=-i\hat{T}^\dagger_\omega \dot{\hat{T}}_\omega\,,
\end{equation}
where the operator $\mathcal H$ is Hermitian and
\begin{equation}
\label{varH}
\text{var}(\mathcal H, \ket{ \psi})=\bra{ \psi}\mathcal H^2\ket{\ \psi}-\bra{\psi}\mathcal H\ket{\psi}^2\,.
\end{equation}

% In the
% general case, the state $\ket{\psi_\omega}$ can be expanded over a
% basis $\ket{\bm n}_c$ of $\omega$-dependent states, with coefficients
% $\psi_{\bm n}(\omega)$ also depending on $\omega$. 

% If the states
% $\ket{\bm n}_c$ are obtained from a basis $\ket{\bm n}_a$ of
% $\omega$-independent states via the unitary transformation
% $\hat{T}_\omega$, one can write $\ket{\psi_\omega}$ as 
% \begin{equation}
% \label{eq:psi}
% \ket{\psi_\omega}=\sum_{\bm n}\psi_{\bm n}(\omega)\hat{T}_\omega\ket{\bm n}_a\,.
% \end{equation}

\subsubsection{Two unitaries and chain rule for the QFI}
\label{chainrule}
Let us now consider the case where the parameter is encoded in $\ket{\psi_\omega}$ by means of two consecutive unitaries depending on $\omega$.
% For example, for the QHE we will see that
% the magnetic field dependence of the single particle
% energy-eigenstates translates into a parameter dependence of the Fock
% states used as basis for the $N$-particle Hilbert space, while
% additional magnetic-field dependence can arise from the shift of the
% energy levels and hence a modification of time-dependent phases in 
% $\psi_{\bm n}(\omega)$. This is quite general:
Our aim is thus to calculate the QFI of a state of the form
\begin{equation}
  \ket{\psi_\omega}=\hat{U}_\omega\hat{T}_\omega\ket{\psi_{\omega_0}}.
\label{eq:psiom} 
\end{equation}
In the same way as $\cH$ in Eq.~\eqref{qfiH}, we define ${\mathcal U}=-i\hat{U}^\dagger_\omega \dot{\hat{U}}_\omega$. From unitarity of $\uom$ and $\tom$ we have
\begin{align}
\cH&=-i\hat{T}^\dagger_\omega \dot{\hat{T}}_\omega=i\dot{\hat{T}}^\dagger_\omega \hat{T}_\omega,\qquad \cH^2=\dot{\hat{T}}^\dagger_\omega \dot{\hat{T}}_\omega\\
\cU&=-i\hat{U}^\dagger_\omega \dot{\hat{U}}_\omega=i\dot{\hat{U}}^\dagger_\omega \hat{U}_\omega,\qquad \cU^2=\dot{\hat{U}}^\dagger_\omega \dot{\hat{U}}_\omega
\end{align}
with $\cH$ and $\cU$ Hermitian. We introduce the state
\begin{equation}
  \ket{\phi_\omega}=\hat{T}_\omega\ket{\psi_{\omega_0}},
  \label{eq:defphi} 
\end{equation}
so that 
\begin{align}
  \ket{\psi_\omega}&= \uom  \ket{\phi_\omega}\\
  \ket*{\dot{\psi}_\omega}&= \dot{\hat{U}}_\omega \ket{\phi_\omega}+\uom  \ket*{\dot{\phi}_\omega},\\
  \ket*{\dot{\phi}_\omega}&=\dot{\hat{T}}_\omega\ket{\psi_{\omega_0}}.
\end{align}
This yields the identities
\begin{align}
\braket*{\dot{\phi}_\omega}&=\expval*{\cH^2}{\psi_{\omega_0}}\\
\braket*{\dot{\phi}_\omega}{\phi_\omega}&=-i\expval*{\cH}{\psi_{\omega_0}}.
\end{align}
From Eq.~\eqref{eq:I} we then obtain
\begin{align}
  \label{eq:QFIchain}
  \frac14I(\ket{\psi_\omega},\omega)&=\text{var}(\cH,\ket{\psi_{\omega_0}})+\text{var}(\cU,\ket{\phi_{\omega}})\\
&-2\Im\mel*{\dot{\phi}_{\omega}}{\cU}{\phi_{\omega}}-2\ev*{\cU}{\phi_{\omega}}\ev*{\cH}{\psi_{\omega_0}}.\nonumber  
\end{align}
Equation \eqref{eq:QFIchain} provides a chain rule for the QFI associated with two unitary operators.
If $\hat{U}_\omega$ or $\hat{T}_\omega$ is the
identity operator, one gets back the expression \eqref{qfiH} for a single operator.  When two
unitaries are present, the variances sum up,  but in addition
there is a cross term that comes from the variation of both $\hat{U}_\omega$
and $\hat{T}_\omega$ with the parameter. Recently, in \cite{PhysRevA.95.012111} a chain
rule was derived for the case that the POVM is a projective
measurement that depends itself on the
parameter.

%*************************************************************************
%*************************************************************************
\section{Some specific cases}\label{specCases}
%*************************************************************************
%*************************************************************************

%*************************************************************************
\subsection{QFI for Bogoliubov transformations}\label{secQFIbog}
%*************************************************************************

We now consider the situation where the parameter $\omega$ is encoded in $\ket{\psi_{\omega}}$ by means of a single unitary transformation $\Twa$ associated with a Bogoliubov transformation. The operator $\Twa$ is defined by Eqs.~\eqref{eq:W}--\eqref{eq:T}, with matrices $W$ and $S$ depending on a parameter $\omega$. In the language of section \ref{single}, particles of type $c$ correspond to parameter value $\omega$ and particles of type $a$ to parameter value $\omega_0$.

This situation is a special case of section \ref{single} where the operator $\hat{S}_\omega$ is quadratic in creation and annihilation operators. The QFI is thus directly given by Eq.~\eqref{qfiH}, where the Hermitian operator ${\mathcal H}$ is ${\mathcal H}=-i\Twa^\dagger \Twad$. Our aim is to reexpress the QFI in terms of the matrices $U$ and $V$ defining the Bogoliubov transformation.

\subsubsection{General case}
Using Eq.~\eqref{eq:debHstand} giving the derivative of an integral, we first rewrite $\mathcal H$ as \cite{Boixo07,JulienThesis2017} 
\begin{eqnarray}
  \label{eq:proof1bis}
  {\mathcal H}&=&\int_0^1 ds \, e^{-is \Swa}\frac{d\Swa}{d\omega}e^{is\Swa}\\
&=&\frac{1}{2}\dot{S}_{ij}\int_0^1ds\,e^{-is\Swa}\alpha_ie^{is\Swa}e^{-is\Swa}\alpha_j
    e^{is\Swa}\,\nonumber
\end{eqnarray}
(again with implicit summation over repeated indices). The
term $e^{-is\Swa}\alpha_ie^{is\Swa}$ can be rewritten as $\Twa^{-s} 
\alpha_i \Twa^s=(\alpha e^{-i s S\Xi})_i$, yielding 
\begin{eqnarray}
  {\mathcal H}&=&\frac{1}{2}\int_0^1 ds\,(\alpha e^{-i
                  sS\Xi})_i\dot{S}_{ij}(\alpha e^{-i sS\Xi})_j\\
&=&\frac{1}{2}\alpha_l \int_0^1 ds\,(e^{-is S
    \Xi})_{li}\dot{S}_{ij}(e^{-is S\Xi })_{kj} \alpha_k \label{eq:proof21}\\
&=&\frac{1}{2}\alpha_l \int_0^1 ds\,(e^{-is S
    \Xi})_{li}\dot{S}_{ij}(e^{is \Xi S})_{jk} \alpha_k\,,  \label{eq:proof2}
\end{eqnarray}
where between \eqref{eq:proof21} and \eqref{eq:proof2} we have used
$(S\Xi)^t=-\Xi S$ due to antisymmetry of $S$. 
The operator ${\mathcal H}$ can thus be expressed as a quadratic form in the $a_i,a_i^\dagger$ as
\begin{equation}
  {\mathcal H}=\frac{1}{2}\alpha\tilde{\Omega}\alpha^t\,,\label{eq:genH}
\end{equation}
with
\begin{equation}
  \label{eq:Omegat}
  \tilde{\Omega}=\int_0^1 ds \,e^{-is S\Xi}\dot{S}e^{is \Xi S}\,.
\end{equation}
In Appendix \ref{app1} we give an alternative proof of \eqref{eq:Omegat} based on the commutation relations of $\hat{S}$ and $\dot{\hat{S}}$. 
The above equation gives the most general expression for the operator whose variance gives the QFI. The remaining integral in Eq.~\eqref{eq:Omegat} makes it uneasy to use. In order to make some progress we now consider a natural additional hypothesis.

\subsubsection{Case $[S,\Xi]=0$}
\label{generatorQFI}
The general result \eqref{eq:genH}--\eqref{eq:Omegat} can be further simplified if we
make the additional assumption that $S$ and $\Xi$ commute:
The block structure \eqref{eq:S} implies that $[S,\Xi]=0$ if and only if
$S=-S^*$, that is, $iS$ is a real matrix.  In such a case, using \eqref{eq:debHstand}, Eq.~\eqref{eq:Omegat} gives
\begin{equation}
   \tilde{\Omega}\Xi=\int_0^1ds \,e^{-is S\Xi}\dot{S}\Xi e^{is S\Xi}=-i W^\dagger \dot{W}\,,
\end{equation}
so that ${\mathcal H}$ becomes
\begin{equation}
  \label{eq:genH2}
  {\mathcal H}=\frac{1}{2}\alpha \Omega \alpha^\dagger\quad\mbox{ with }\quad
  \Omega=-i W^\dagger \dot{W}\,
\end{equation}
(we used the identity $\Xi\alpha^t=\alpha^\dagger$ mentioned below Eq.~\eqref{eq:T}).
Thus, in such a case where $iS$ is real, the matrix $W$ that defines the Bogoliubov
transformation, together with its derivative with respect to the parameter
$\omega$, provide an expression for $\mathcal H$ as a quadratic
form of the operators $a_i^\dagger, a_i$. 

Below, we will be interested in the calculation of the QFI as a function of $\omega$ in the vicinity of a fixed parameter $\omega_0$. We will therefore
evaluate all quantities in the limit $\omega\to \omega_0$. In this
limit, the Bogoliubov transformation goes to the identity, so that
we have $U\to \text{Id}_M$ and $V\to 0$. The matrix $\dot{W}$ then
involves derivatives of $U$ and $V$ with respect to $\omega$ taken at
$\omega\to \omega_0$, that we will denote $\dot{U}$ and
$\dot{V}$. 
Taking the derivative of the relations $U^\dagger U+V^\dagger
V=\text{Id}_M$ and $U^tV+V^tU=0_M$ with respect to $\omega$ and 
% taking
then
the limit $\omega\to \omega_0$ we get $\dot{U}+\dot{U}^t=0$ and
$\dot{V}+\dot{V}^t=0$; this can be shown by using the fact that since $iS$ is real
then $W=\exponential(i S \Xi)$ is, too (with $\Xi$ defined by Eq.~\eqref{eq:X}), and therefore also $U$ and $V$.
With this antisymmetry of
$\dot{U}$ and $\dot{V}$ together with the fermionic anticommutation
relations, Eq.~\eqref{eq:genH2} becomes, using antisymmetry of $\dot{U}$ and $\dot{V}$,
\begin{equation}
  \label{eq:genH2bp}
 {\mathcal H}=-i\sum_{k<l}\left(\dot{U}_{kl}a_k^\dagger a_l-\dot{U}_{kl}a_l^\dagger a_k+\dot{V}_{kl}a_k^\dagger a_l^\dagger+\dot{V}_{kl}a_k a_l\right).
\end{equation}

\subsubsection{Case $U=R$  real and $V=0$}
{From now on we will specialize to the case where the Bogoliubov transformation does not mix creation and annihilation operators, i.e.~$V=0$, and the unitary transformation $U$ is orthogonal. This case is of great relevance, since it is precisely the framework in which we will derive expressions in Section \ref{sec:qhe}. Indeed, in the context of the quantum Hall effect the Bogoliubov transformation is given by the matrix $R$ of overlaps \eqref{defR}, whose entries are real. The fact that these overlaps are real is a consequence of the structure of the Hall wavefunctions in the Landau gauge, given by Eq.~\eqref{eq:WF} below: the complex phase is a plane wave that does not depend on the parameter $\omega$, yielding a (real) delta function in the overlap.}
   
In this case, $W$ is real and one can always choose a matrix $S$ in \eqref{eq:Sdef} such that $iS\Xi$ is real. Hence $iS$ is real,
and as a consequence $[S,\Xi]=0$. Equation \eqref{eq:genH2bp} can thus be used, and it gives
\begin{equation}
  \label{eq:genfin}
  {\mathcal H}=-i \sum_{k<l}\dot{R}_{kl}(a_k^\dagger a_l-a_l^\dagger a_k).
\end{equation}

\subsubsection{QFI for a basis state}
\label{urv0}
Let us consider the case where $\ket{\psi_\omega}$ is the parameter-dependent basis state $\ket{\bm n}_{\omega}=\hat{T}_{\omega}\ket{\bm n}_{\omega_0}$. Again we associate mode $c$ with $\omega$ and mode $a$ with $\omega_0$.
According to \eqref{qfiH}, the QFI is given by the variance $I(\ket{\bm n}_{\omega},\omega)=4\text{var}(\mathcal H, \ket{\bm n}_{\omega_0})$.

In the remainder of the paper we will only address the case where $[S,\Xi]=0$, in which case $\mathcal H$ is given by the expression \eqref{eq:genH2bp}. It only requires to calculate $\mathcal H\ket{\bm n}_a$. We have
\begin{align}
a_k\ket{\bm n}_a&=\delta_{n_k,1}\,(-1)^{\sum_{j=0}^{k-1}n_j}\ket{\overline{\bm n}^{{}_{k}}}_a,\\
a_k^\dagger\ket{\bm n}_a&=\delta_{n_k,0}\,(-1)^{\sum_{j=0}^{k-1}n_j}\ket{\overline{\bm n}^{{}_{k}}}_a,
  \label{eq:akakdag}
\end{align}
where $\ket{\overline{\bm n}^{{}_{k}}}_a$ is the state $\ket{\bm n}_a$
with $n_k$ replaced by $1-n_k$ (i.e.~the $k$th ``bit'' in the binary string
$\bm n$ is  flipped).
This leads (for $k<l)$ to 
\begin{align}
 \label{eq:ak0al1}
a_k^\dagger a_l\ket{\bm n}_a&=\delta_{n_k,0}\delta_{n_l,1}\,(-1)^{\sum_{j=k}^{l-1}n_j}\ket{\overline{\bm n}^{{}_{k,l}}}_a,\\
a_l^\dagger a_k\ket{\bm n}_a&=-\delta_{n_k,1}\delta_{n_l,0}\,(-1)^{\sum_{j=k}^{l-1}n_j}\ket{\overline{\bm n}^{{}_{k,l}}}_a,
\end{align}
where in $\ket{\overline{\bm n}^{{}_{k,l}}}_a$ bits $k$ and $l$ are flipped. Similarly we have (still for $k<l$)
\begin{align}
 \label{eq:ak1al1}
a_k^\dagger a_l^\dagger\ket{\bm n}_a&=\delta_{n_k,0}\delta_{n_l,0}\,(-1)^{\sum_{j=k}^{l-1}n_j}\ket{\overline{\bm n}^{{}_{k,l}}}_a,\\
a_k a_l\ket{\bm n}_a&=\delta_{n_k,1}\delta_{n_l,1}\,(-1)^{\sum_{j=k}^{l-1}n_j}\ket{\overline{\bm n}^{{}_{k,l}}}_a.
\end{align}
Inserting these expressions into Eq.~\eqref{eq:genH2bp} leads to
\begin{align}
  \label{hpsi}
  {\mathcal H}\ket{\bm n}_a&=-i\sum_{k<l}\left(\dot{U}_{kl}\delta_{n_k,0}\delta_{n_l,1}+\dot{U}_{kl}\delta_{n_k,1}\delta_{n_l,0}\right.\nonumber\\
  &\left.+\dot{V}_{kl}\delta_{n_k,0}\delta_{n_l,0}+\dot{V}_{kl}\delta_{n_k,1}\delta_{n_l,1}\right)\,(-1)^{\sum_{j=k}^{l-1}n_j}\ket{\overline{\bm n}^{{}_{k,l}}}_a,
\end{align}
which can be shortened to
\begin{equation}
  \label{hpsi2}
  {\mathcal H}\ket{\bm n}_a=-i\sum_{k<l}(-1)^{\sum_{j=k}^{l-1}n_j}\,D^{(|n_k-n_l|)}_{kl}\ket{\overline{\bm n}^{{}_{k,l}}}_a,
\end{equation}
with $D^{(0)}=\dot{V}$ and $D^{(1)}=\dot{U}$.
Since each flipped state $\ket{\overline{\bm n}^{{}_{k,l}}}_a$ is orthogonal to $\ket{\bm n}_a$, we have ${}_a\bra{\bm n}\mathcal H\ket{\bm n}_a=0$. The quadratic term in \eqref{varH} is given by the square of the 2-norm of $\mathcal H\ket{\bm n}_a$. Since all terms in the sum \eqref{hpsi2} are orthogonal to each other, the QFI finally reads
\begin{equation}
\label{Ibasisstate}
I(\ket{\bm n}_\omega,\omega)=4\sum_{k<l}|D^{(|n_k-n_l|)}_{kl}|^2\,.
\end{equation}
In the case where $U=R$ is real and $V=0$, ${\mathcal H}$ is given by Eq.~\eqref{eq:genfin}. Only $D^{(1)}=\dot{R}$ contributes, so that Eq.~\eqref{Ibasisstate} reduces to
\begin{equation}
\label{Ibasisstate2}
I(\ket{\bm n}_\omega,\omega)=4\sum_{\genfrac{}{}{0pt}{}{k<l}{|n_k-n_l|=1}}|\dot{R}_{kl}|^2\,.
\end{equation}
This is the expression which we shall use in Section \ref{sec:qhe} in the context of the quantum Hall effect.

It is interesting to analyze Eq.~\eqref{Ibasisstate2} in the context of a finite-dimensional Hilbert space. The sum in \eqref{Ibasisstate2} is a sum over all pairs of occupied and unoccupied modes. For a finite-dimensional Hilbert space of single-particle states where each
state is occupied (e.g.~an insulating band in a solid), this sum vanishes. Indeed, as the corresponding Fock space is one-dimensional, all parameter dependence
through unitary transformations amongst the annihilators trivially reduces to a phase, which cancels in the density matrix.  Hence, the state is independent of the
parameter under such unitaries, as can be checked explicitly for $N=2$, which is consistent with the fact that the QFI is zero. This implies of course, that $\omega$
cannot be measured at all, but not that the variance of any unbiased
estimator diverges.  Rather, the conditions for the QCRB break down:
one cannot have an unbiased estimator in an $\epsilon$-interval about
the 
true value $\omega$ if the state is independent of $\omega$:
$\langle\hat{\omega}\rangle=\omega$ can only be true at a single point
if the lhs is independent of $\omega$, not in a whole finite interval,
even if it is arbitrarily small.

\subsection{QFI for a Hamiltonian evolution} 
\label{twounitexample}
In Section \ref{chainrule} we obtained the QFI associated with a state obtained by applying an operator $\hat{T}_\omega$ followed by an operator $\hat{U}_\omega$. It is expressed via the chain rule \eqref{eq:QFIchain}. This expression takes a much simpler form in the case where $\hat{U}_\omega$ is the evolution operator associated with the $\omega$-dependent Hamiltonian
\begin{equation}
 \hat{H}_\omega=\sum_k\epsilon_k(\omega)\hat{n}_k(\omega)\,,
\end{equation}
describing a system of non-interacting fermions. Here $\epsilon_k(\omega)$ are the
parameter-dependent single-particle energy eigenvalues, and $\hat{n}_k$ are the occupation number operators. 

Let the initial state be parameter-independent (or, equivalently, taken at a fixed value $\omega_0$ of the parameter). Starting at time $t=0$ from $\ket{\psi_{\omega_0}(0)}=\sum_\bn \psi_\bn\ket{\bn}_{\omega_0}$, the state at time $t$ is most easily calculated by first changing the basis $\ket{\bm n}_{\omega_0}$ to $\ket{\bm n}_\omega$ by means of a Bogoliubov transformation, as in Section \ref{bogtra}: particles of type $c$ correspond to parameter value $\omega$ and particles of type $a$ to parameter value $\omega_0$. We then have, from Eq.~\eqref{eq:proofpsi}, the identity  
\begin{equation}
\label{nn0}
 \ket{\bm n}_{\omega}=\hat{T}_{\omega}\ket{\bm n}_{\omega_0}\,.
\end{equation}
The evolution operator $\hat{U}_\omega= \exponential(-i\hat{H}_\omega t)$ is diagonal in the basis $\ket{\bm n}_\omega$, so that
\begin{align}
  \label{eq:psit2}
    \ket{\psi_\omega(t)}&=\hat{U}_\omega\hat{T}_{\omega}\ket{\psi_{\omega_0}(0)}\nonumber\\
&=e^{-i\hat{H}_\omega t}\sum_{\bm n}\psi_{\bm n}\ket{\bm n}_\omega \nonumber\\
&=\sum_{\bm n}\psi_{\bm n}e^{-iE_\bn(\omega) t}\ket{\bm n}_{\omega}\nonumber\\
&=\sum_{\bm n}\psi_{\bm  n}(\omega,t)\ket{\bm n}_{\omega}\,,
\end{align}
with $\psi_{\bm n}(\omega,t)=\psi_\bn e^{-i E_{\bm n}(\omega)t}$ and $E_{\bm n}(\omega)=\sum_k\epsilon_k(\omega) n_k$ the total energy of many-body basis state $\ket{\bm n}_\omega$. Thus one can go from $\ket{\psi_{\omega_0}(0)}$ to a state of the form \eqref{eq:psigeneral} with two unitaries, one for the change of basis and the other for time evolution.

Our aim is to calculate the QFI of
\begin{equation}
   \ket{\psi_\omega}=\sum_{\bm n}e^{-iE_\bn(\omega) t}\psi_{\bm n}\ket{\bm n}_{\omega},
\end{equation}
where $\psi_{\bm n}$ are the coordinates of the initial state  $\ket{\psi_{\omega_0}(0)}$ in the basis $\ket{\bn}_{\omega_0}$ and thus are
independent of $\omega$. We introduce 
 \begin{align}
 \label{defgamma}
   \ket{\gamma_\omega}&= \sum_{\bm n}\left(-i\dot{E}_\bn(\omega)t\right)e^{-iE_\bn(\omega) t}\psi_{\bm n}\ket{\bm n}_{\omega},\\
   \ket{\chi_\omega}&= \sum_{\bm n}e^{-iE_\bn(\omega) t}\psi_{\bm n}\ket{\dot{\bm n}}_{\omega},\\
   \ket{\varphi_{\omega}}&= \sum_{\bm n}e^{-iE_\bn(\omega) t}\psi_{\bm n}\ket{\bm n}_{\omega_0},
   \label{defchi}
   \end{align}
so that $\ket*{\dot{\psi}_\omega}=   \ket{\gamma_\omega}+ \ket{\chi_\omega}$. In terms of $\ket{\gamma_\omega}$ and $\ket{\chi_\omega}$, the QFI  Eq.~\eqref{eq:I}  reads
 \begin{align}
  \label{eq:QFIUT}
  \frac14I(\ket{\psi_\omega},\omega)&=\braket{\gamma_\omega}+\braket{\gamma_\omega}{\psi_\omega}^2
  +\braket{\chi_\omega}+\braket{\chi_\omega}{\psi_\omega}^2\nonumber\\
  &+\braket{\gamma_\omega}{\chi_\omega}+\braket{\chi_\omega}{\gamma_\omega}+2\braket{\gamma_\omega}{\psi_\omega}\braket{\chi_\omega}{\psi_\omega}.
\end{align}
One then readily gets from \eqref{defgamma}
\begin{equation}
\label{gaga}
 \braket{\gamma_\omega}+\braket{\gamma_\omega}{\psi_\omega}^2
 =\sum_\bn
|\psi_\bn|^2\dot{E}_\bn^2t^2-(\sum_\bn |\psi_\bn|^2\dot{E}_\bn t)^2.
\end{equation}
If we define the diagonal operator $\dot{E}=\sum_\bn \dot{E}_\bn\ketbra{\bm n}_{\omega_0}$, this gives
\begin{equation}
\label{gaga2}
 \braket{\gamma_\omega}+\braket{\gamma_\omega}{\psi_\omega}^2
 =\text{var}(\dot{E} t,\varphi_{\omega}).
\end{equation}
Deriving \eqref{nn0} with respect to $\omega$, we get
\begin{equation}
\label{nn0dot}
\ket{\dot{\bm n}}_{\omega}= \dot{\hat{T}}_\omega\hat{T}^\dagger_\omega\ket{\bm n}_{\omega}
\end{equation}
and thus
 \begin{equation}
 \ket*{\chi_\omega}=  \dot{\hat{T}}_\omega\hat{T}^\dagger_\omega  \ket{\psi_\omega}=  \dot{\hat{T}}_\omega\ket{\varphi_{\omega}}.
    \end{equation}
This yields $\braket{\chi_\omega}{\psi_\omega}=-i\expval*{\cH}{\varphi_{\omega}}$ and thus
\begin{equation}
  \label{chichi}
  \braket{\chi_\omega}+\braket{\chi_\omega}{\psi_\omega}^2=\text{var}(\cH, \varphi_{\omega}).
  \end{equation}
Noting that
\begin{equation}
\ket{\dot{\varphi}_{\omega}}= \sum_{\bm n}\left(-i\dot{E}_\bn(\omega)t\right)e^{-iE_\bn(\omega) t}\psi_{\bm n}\ket{\bm n}_{\omega_0}=\hat{T}^\dagger_\omega\ket{\gamma_\omega},
  \end{equation}
the last contribution in Eq.~\eqref{eq:QFIUT} involves the terms
\begin{equation}
\braket{\gamma_\omega}{\psi_\omega}=i\sum_\bn |\psi_\bn|^2\dot{E}_\bn t\equiv i\expval*{\dot{E} t}_{\psi_{\omega_0}}=i\expval*{\dot{E} t}_{\varphi_{\omega}}
  \end{equation}
and
  \begin{equation}
\braket{\gamma_\omega}{\chi_\omega}=i\mel*{\dot{\varphi}_{\omega}}{\cH}{\varphi_{\omega}}=-\expval*{\dot{E} t\,\, \cH}_{\varphi_{\omega}}.
  \end{equation}
We obtain
\begin{align}
\braket{\gamma_\omega}{\chi_\omega}+\braket{\chi_\omega}{\gamma_\omega}&=-\expval*{\dot{E} t\,\, \cH}_{\varphi_{\omega}}-\expval*{\cH\,\,\dot{E} t}_{\varphi_{\omega}}\\
 \braket{\gamma_\omega}{\psi_\omega}\braket{\chi_\omega}{\psi_\omega}&=2\expval*{\dot{E} t}_{\varphi_{\omega}}\expval*{\cH}_{\varphi_{\omega}}.
\end{align}
Summing together all contribution in \eqref{eq:QFIUT} we get
\begin{align}
    \label{eq:QFIUTfin}
 \frac14I(\ket{\psi_\omega},\omega)&=\text{var}(\dot{E} t,\varphi_{\omega})+\text{var}(\cH, \varphi_{\omega})\\
  &-\expval*{\dot{E} t\,\, \cH}_{\varphi_{\omega}}-\expval*{\cH\,\,\dot{E} t}_{\varphi_{\omega}}+2\expval*{\dot{E} t}_{\varphi_{\omega}}\expval*{\cH}_{\varphi_{\omega}},\nonumber\\
 &= \expval*{(\dot{E} t-\cH)^2}_{\varphi_{\omega}}-\expval*{\dot{E} t-\cH}^2_{\varphi_{\omega}}.
 \end{align}
We thus obtain the very compact expression
\begin{equation}
  \label{eq:QFIUTfin2}
 I(\ket{\psi_\omega},\omega)=4\,\text{var}(\dot{E} t-\cH, \varphi_{\omega}),
  \end{equation}
  with  $\dot{E}=\sum_\bn \dot{E}_\bn\ketbra{\bm n}_{\omega_0}$ and 
  \begin{align}
 {}_{\omega_0}\!\!\bra{\bm m}\cH\ket{\bm n}_{\omega_0}=i\,\,{}_{\omega_0}\!\!\bra{\bm m}\dot{\hat{T}}^\dagger_\omega \hat{T}_\omega\ket{\bm n}_{\omega_0}=i\,\,{}_{\omega}\!\!\bra{\dot{\bm m}}\ket{\bm n}_{\omega}.
  \end{align}
 At $t=0$, state $\ket{\varphi_\omega}$ coincides with $\ket{\psi_{\omega_0}}$ and thus we recover the QFI for a single unitary, Eq.~\eqref{qfiH}.

\subsection{QFI for a general state}
We now put together the results from the previous two subsections and consider the case of a superposition $\ket{\psi_\omega}=\sum_{\bm n} \psi_{\bm n}(\omega)\ket{\bm n}_\omega$ of basis states. The QFI is given by Eq.~\eqref{eq:QFIUTfin2}, that is, by the variance of $\dot{E}t-{\mathcal H}$ in state $\ket{\varphi_\omega}$. That state is defined by \eqref{defchi}, {namely, 
  \begin{equation}
    \label{defvarphibis}
    \ket{\varphi_{\omega}}= \sum_{\bm n}e^{-iE_\bn(\omega) t}\psi_{\bm n}\ket{\bm n}_{\omega_0}
  \end{equation}
  in the basis of kets $\ket{\bm n}_{\omega_0}$.
Operator ${\mathcal H}$ corresponds to a Bogoliubov transformation and its action on basis states $\ket{\bm n}_{a}$ is given by Eq.~\eqref{hpsi2}. Since $\omega_0$ is the frequency for type-$a$ particles, we have
\begin{equation}
  \label{hpsi2bis}
  {\mathcal H}\ket{\bm n}_{\omega_0}=-i\sum_{k<l}(-1)^{\sum_{j=k}^{l-1}n_j}\,D^{(|n_k-n_l|)}_{kl}\ket{\overline{\bm n}^{{}_{k,l}}}_{\omega_0}.
\end{equation}
%We have
%\begin{align}
%  \text{var}(\dot{E}t,\varphi_{\omega})=\sum_{\bm n}(\dot{E}_{\bm n}t)^2|\psi_{\bm n}|^2-
%  \left(\sum_{\bm n}\dot{E}_{\bm n}t|\psi_{\bm n}|^2\right)\,.
%\end{align}
By linearity, Eqs.~\eqref{defvarphibis}-\eqref{hpsi2bis} directly give}
\begin{equation}
  \label{hpsi3}
  {\mathcal H}\ket{\varphi_{\omega}}=-i\sum_{\bm n}\psi_{\bm n}(\omega,t)\sum_{k<l}(-1)^{\sum_{j=k}^{l-1}n_j}\,D^{(|n_k-n_l|)}_{kl}\ket{\overline{\bm n}^{{}_{k,l}}}_{\omega_0}\,,
\end{equation}
with $\psi_{\bm n}(\omega,t)=\psi_{\bm n}e^{-iE_\bn(\omega) t}$
Permuting the two sums, we make the change from $\overline{\bm n}^{{}_{k,l}}$ to ${\bm n}$ in the sum over ${\bm n}$. This does not change the term $D^{(|n_k-n_l|)}_{kl}$, while flipping $n_k$ changes the overall sign. This leads to
\begin{align}
  {\mathcal H}\ket{\varphi_{\omega}}&=\sum_{\bm n}h_{\bm n}(\omega)\ket{\bm n}_{\omega_0}\,,\nonumber\\
h_{\bm n}(\omega)&=\sum_{k<l}(-1)^{\sum_{j=k}^{l-1}n_j}\,D^{(|n_k-n_l|)}_{kl}\psi_{\overline{\bm n}^{{}c_{k,l}}}(\omega,t)\,.
  \label{hpsi4}
\end{align}
The vectors ${\mathcal H}\ket{\varphi_{\omega}}$ are now both expressed in the same basis $\ket{\bm n}_{\omega_0}$, so that
\begin{equation}
 \left(\dot{E}t- {\mathcal H}\right)\ket{\varphi_{\omega}}=\sum_{\bm n}\left(\dot{E}_{\bm n}t\psi_{\bm n}(\omega,t)-h_{\bm n}(\omega)\right)\ket{\bm n}_{\omega_0}\,.
  \label{eth}
\end{equation}
We therefore get
\begin{align}
  \text{var}(\dot{E}t-\mathcal H, \varphi_\omega)&= \sum_{\bm n}\left|\dot{E}_{\bm n}t\psi_{\bm n}(\omega,t)-h_{\bm n}(\omega)\right|^2\nonumber\\
  &-
  \left( \sum_{\bm n}(\dot{E}_{\bm n}t|\psi_{\bm n}|^2-\psi_{\bm n}^*(\omega,t)h_{\bm n}(\omega))\right)^2\,.\label{varFin}
\end{align}
For a single particle the calculation can be done more easily starting directly from \eqref{eq:I}. One checks that in that case one gets \eqref{varFin} with $\bm n$ replaced by the index of the single particle states.

%*************************************************************************
%*************************************************************************
\section{Application to Quantum Hall effect}
%*************************************************************************
%*************************************************************************
\label{sec:qhe}

%*************************************************************************
\subsection{Single-particle quantum Hall physics}
%*************************************************************************
We now turn to an application of our results to quantum Hall
physics. We consider a two-dimensional system of size $L$ along the
$x$-axis and $w$ along the $y$-axis, subjected to a perpendicular
constant magnetic field $B$ along the $z$-axis.
We choose the coordinate system such that $|y|\le w/2$, $0\le x \le L$.
We denote by ${\cal A}=L w$ the area of the sample. The frequency $\omega=eB/m_{\rm eff}$ is the cyclotron frequency of charge
carriers with effective mass $m_{\rm eff}$, $l_B=\sqrt{\hbar/(eB)}=\sqrt{\hbar/(m_{\rm eff}\omega)}$ is the magnetic length, and at the same time the oscillator length associated with frequency $\omega$.  We denote with $n_B=1/(2\pi l_B^2)=B/(h/e)$ the magnetic flux density (number of flux quanta  
$\Phi_0=h/e$ 
per unit area), and $M =n_B {\cal A}$ is the total number of flux quanta.% in the sample that hence determines the degeneracy of each Landau level (LL).

In the Landau gauge $\bm A=(-B y,0,0)$, one can make the Ansatz that the wave function factorizes in $x$ and $y$ direction. Choosing periodic boundary conditions in the $x$-direction results in  plane waves in $x$ with wave vector of the form $k_m=m(2\pi/L)$. The effective total
Hamiltonian is then given by
\begin{equation}
  \label{eq:H}
  H=\frac{p_y^2}{2m_{\rm eff}}+\frac{1}{2}m_{\rm eff}\omega^2(y-y_m)^2\,,
\end{equation}
where  $y_m=k_m l_B^2$ is a shift of the oscillator in the $y$ direction that depends on the quantum number $m$ of the quantization in $x$-direction. The kinetic energy of the plane wave is contained in the $y_m^2$ term. As a consequence, $m$ enters only through the shift
$y_m$ in \eqref{eq:H} of the origin of the oscillator, and thus energy eigenvalues do not depend on $m$: Landau levels are degenerate. The condition
$|y_m|\le w/2$ is equivalent to $|m|\le {\cal A}/(4\pi l_B^2)={\cal A}
m_{\rm eff}\omega/(4\pi\hbar)=\Phi/(2\Phi_0)$, which for $\omega$ the
cyclotron frequency amounts to $|m|\le M/2$.  This is the well-known
result that the number $M$ of states per Landau level (LL) $n$, and hence degeneracy of each energy eigenvalue $\hbar\omega_\text{eff} n$, is given by the number of flux quanta
 through the surface.   For simplicity we will assume $M$ to be odd, so that $m$ takes the values $m=-(M-1)/2,\ldots,0,\ldots (M-1)/2$.

The energy eigenstates $|n,m\rangle_{\omega}$ can be labeled with the
two quantum numbers $n,m$. They are conveniently described   
in the chosen Landau gauge by the wave functions 
\begin{equation}\label{eq:WF} 
\langle x,y|n,m\rangle_{\omega}=\frac{e^{i k_m x}}{\sqrt{L}}\chi_{n,m}\left(\frac{y-k_ml_B^2}{l_B}\right),
\end{equation}
where 
\begin{equation}\label{eq:ho}
 \chi_{n,m}(\eta)=\frac{{\cal N}_n}{\sqrt{l_B}}H_n(\eta)e^{-\eta^2/2}
\end{equation}
is the usual harmonic-oscillator wave function in terms of the Hermite polynomial $H_n(\eta)$, while 
\begin{equation}
\label{Nn}
 {\cal N}_n=\frac{1}{\sqrt{2^n \sqrt{\pi}n!}}
\end{equation}
is a normalization factor.

%*************************************************************************
\subsection{Wave function overlaps}
%*************************************************************************
In order to calculate the QFI using Eq.~\eqref{Ibasisstate2}, we first need to obtain the derivative of the matrix $R$ of overlaps.
We calculate the overlap between states $|n,m\rangle_{\omega_0+\delta\omega}$, where the frequency differs from $\omega_0$ by an infinitesimal amount $\delta\omega$, and states $|n',m'\rangle_{\omega_0}$ at some fixed frequency $\omega_0$. At first order, 
\begin{align}
\label{eq:overlapom}
 _{\omega_0}\langle n',m'|n,m\rangle_{\omega_0+\delta\omega}&\simeq \delta_{n,n'}\delta_{m,m'}\nonumber\\
 & +\ _{\omega_0}\langle n',m'|\partial_{\omega_0}|n,m\rangle_{\omega_0}\delta\omega.
\end{align}
The calculation of the first-order term is detailed in Appendix \ref{calculfinaloverlap}. We get the final expression
\begin{widetext}
{
\begin{align}
  \nonumber
  {}_{\omega_0}\langle n',m'|n,m\rangle_{\omega_0+\delta\omega}&\simeq \delta_{n,n'}\delta_{m,m'}
                                                                 +\left[\frac{k_m l_B}{\sqrt{2}\omega_0}\left(\sqrt{n}\delta_{n',n-1}-\sqrt{n+1}\delta_{n',n+1}\right)\right.\\
  &\left.+\frac{1}{4\omega_0} \left(\sqrt{n(n-1)}\delta_{n',n-2}-\sqrt{(n+2)(n+1)}\delta_{n',n+2} \right)\right]\delta_{m',m}\delta\omega.
 \label{finaloverlap}
\end{align}
An alternative way of deriving this quantity is to start from the Hutchisson formula \cite{smith_overlap_1969} for $R_{n'n}={_{\omega_0}}\braket{n'}{n}_\omega$, given by
\begin{align}
\label{Hutch}R_{n'n}=\sqrt{2^{-(n+n')}qn!n'!}(-1)^{n}e^{-\frac14\gamma^2p}\sum_{r=0}^{\min(n,n')}
\frac{(-2q)^r}{r!}
\sum_{s=0}^{\lfloor (n-r)/2\rfloor}\frac{(\gamma p)^{n-r-2s}}{(n-r-2s)!}\frac{x^s}{s!}\sum_{t=0}^{\lfloor (n'-r)/2\rfloor}\frac{(\gamma q)^{n'-r-2t}}{(n'-r-2t)!}\frac{(-x)^t}{t!},
\end{align}
with $\gamma=k_m(l_B^2(\omega)-l_B^2(\omega_0))/l_B(\omega)$,
$x=(\omega-\omega_0)/(\omega+\omega_0)$,
$q=2(\omega\omega_0)^{1/2}/(\omega+\omega_0)$, and
$p=2\omega_0/(\omega+\omega_0)$ (we took the formula of \cite{smith_overlap_1969} with $\nu'=\omega_0$ and $\nu''=\omega$). This approach is more cumbersome. As a check, we show in Appendix \ref{otherproofoverlap} that a first-order expansion of \eqref{Hutch} around $\omega=\omega_0$ allows us to recover the result \eqref{finaloverlap}.\\
}
\end{widetext}

%************************************************************************************************
\subsection{Gauge choice and occupation numbers}
%************************************************************************************************
We now consider the basis state $\ket{\bm n}_\omega$ where $N$
particles fill the lowest available energy levels.
The QFI is given by  \eqref{Ibasisstate2}, which involves a sum over all pairs of labels $(k,l)$ with $k<l$, so that only labels such that the occupation number differs by 1 contribute. That is, since only the lowest levels are filled, the sum runs over all pairs $k<l$ such that level $k$ is occupied and level $l$ is empty. In the present context of the quantum Hall effect, each label $k$ has to be replaced by two quantum numbers $(n,m)$; the QFI is thus a sum over pairs of contributions such that level $(n,m)$ is full and level $(n',m')$ is empty.

The filling factor $\nu=N/M$ determines how many LLs are occupied.
The largest integer smaller than $\nu$ is denoted by $f$. In an
infinitely extended sample without additional potentials (``ideal
sample''), it determines the last fully occupied LL. The
last LL is occupied  by only $\mbar$ particles, with $N=M f+\mbar$.

In an ideal
sample all single-electron states with the same $n$ are degenerate in
energy, and the larger the value of $k_m$ the larger
the sensitivity of these states to a change of magnetic field. Indeed $k_m$
determines how quickly the wavefunctions oscillate, and hence how
sensitive they are to a change of $l_B$ with $B$. Importantly, while the different values
that $k_m$ can take are to a certain degree arbitrary, as they depend on the choice of
the basis for the plane waves, the absolute values of $k_m$ matter for the QFI, as the latter
will depend on the energy eigenstate considered.

However, in a real sample, the degeneracy in energy is broken by the
confining potential. The order in which LLs are occupied is nontrivial,
and this can influence the QFI. Typically, in a sensor based on a two-dimensional electron
gas (2DEG), confined electrostatically by metallic electrodes at a
substantial distance from the 2DEG, typically of the order of 100\,nm, %\qq
the confining potential varies on a length scale typically much larger
than $l_B$ for a magnetic field of the order of $1T$. The additional
potential hardly modifies the electron wave functions in this case and
hence just leads to a shift of the energy eigenvalues corresponding to
the value of the potential 
where the energy eigenstate is localized.
By symmetry, one can expect the minimum of the confining potential to lie at the center of
the sample, where it can be approximated by a slow-varying potential.
The shorter the sample in a given direction,
the stronger a variation of the confining potential in that direction, and hence,
the larger the additional potential energy. This implies that the lowest-energy
single electron states to be populated are oscillator states extended
in the {\em largest} direction of the sample,
% starting from $k_m=0$ and then increasing symmetrically with alternating signs,
where the potential energy due to the confining potential grows more slowly.
% with $|m|$ than if the oscillator states were extended along the short direction.
This dictates a
corresponding choice of the Landau gauge: $A_x=-B y$ for $w\gg L$, but
$A_y=B x$ for $L\gg w$ (with the other components of $\bm A$ equal
zero). For $w\simeq L$, a symmetric Landau gauge $\bm A=B(-y,x,0)$ is
most appropriate and leads to axial-symmetric wave functions and
conservation of angular momentum instead of linear momentum.  But
since the additional potential is, for $w=L$, also symmetric under
$w\leftrightarrow L$, both $A_x=-B y$ and $A_y=B x$ should lead to the
same result as the symmetric gauge in this case. \\

In addition to this confining potential, there is typically also a
disorder potential in a real sample.  Disorder arises from impurities
or dopants that are in general relatively far from the 2DEG as
well, and hence lead to a random potential that varies slowly over the
sample.  Energy eigenstates are then localized at the minimum of this
potential and filled in order of increasing energy, like puddles. The
quantum number $m$ ceases to be a good quantum number and is replaced
by a quantum number that labels the position where the oscillator
state corresponding to the Landau levels
are localized.  This implies a QFI that varies randomly from
sample to sample, with a statistics that is, however, beyond the scope
of the paper.  
In the following we restrict ourselves to a clean
sample with only a confining potential that breaks the degeneracy in
energy of the LLs.

%************************************************************************************************
\subsection{Quantum Fisher information for the $N$-particle quantum Hall effect}
%************************************************************************************************
Focusing on the case $w\gg L$ with $A_x=-B y$, the QFI \eqref{Ibasisstate2} can  be expressed as  
  \begin{align}
  \label{eq:Ifinal2b}
I(\ket{\bm n}_\omega,\omega)&=  4(\sum_{m=-\frac{M-1}{2}}^{\mbar-\frac{M+1}{2}}\sum_{n=0}^f\sum_{n'=f+1}^\infty \nonumber\\
&+\sum_{m=\mbar-\frac{M-1}{2}}^{\frac{M-1}{2}}\sum_{n=0}^{f-1}\sum_{n'=f}^\infty )|\dot{R}_{n',m;n,m}|^2\,,
  \end{align}
  with
\begin{align}
\label{Rdot}
&\dot{R}_{n',m';n,m}=\left[\frac{k_m l_B \left(\sqrt{n}\delta_{n',n-1}-\sqrt{n+1}\delta_{n',n+1}\right)}{\sqrt{2}\omega_0}\right.\nonumber\\
&\left.+\frac{\sqrt{n(n-1)}\delta_{n',n-2}-\sqrt{(n+2)(n+1)}\delta_{n',n+2} }{4\omega_0} \right]\delta_{m'm}
\end{align}
the derivative of the overlap between a level $(n,m)$ and a level $(n',m')$, obtained from Eq.~\eqref{finaloverlap}. Note that from the delta function in \eqref{Rdot} only pairs with $m=m'$ contribute.
Only terms with $n<n'$ contribute to the sum, so that only terms $\delta_{n',n+1}$ and $\delta_{n',n+2}$ survive. Equation \eqref{eq:Ifinal2b} reduces to 
  \begin{align}
  \label{eq:Ifinal3}
&I(\ket{\bm n}_\omega,\omega)= 4\left(\sum_{m=-\frac{M-1}{2}}^{\mbar-\frac{M+1}{2}}\sum_{n=0}^f\sum_{n'=f+1}^\infty +\sum_{m=\mbar-\frac{M-1}{2}}^{\frac{M-1}{2}}\sum_{n=0}^{f-1}\sum_{n'=f}^\infty\right)\nonumber\\
&\times\left( \frac{(k_m l_B )^2(n+1)}{2\omega^2}\delta_{n',n+1}+\frac{(n+2)(n+1)}{16\omega^2}\delta_{n',n+2}\right)\,,
  \end{align}
where we have set $\omega_0=\omega$.  
Only pairs $n,n'$ differing by 1 or 2 units contribute, thus \eqref{eq:Ifinal3} becomes
 \begin{align}
  \label{eq:Ifinal4}
&I(\ket{\bm n}_\omega,\omega)=\nonumber\\
&\frac{1}{\omega^2}\left[\sum_{m=-(M-1)/2}^{\mbar-(M+1)/2}
\left( 2(k_m l_B )^2(f+1)+\frac{1}{2}(f+1)^2\right)\right.
\nonumber\\
&\left.+\sum_{m=\mbar-(M-1)/2}^{(M-1)/2}
\left( 2(k_m l_B )^2f+\frac{1}{2}f^2\right)\right].
\end{align}
Replacing $k_m$ by its value $2\pi m/L$, one can perform the sum over
$m$. This  leads to the final expression
\begin{align}
  \label{eq:Ifinal5}
  I(\ket{\bm n}_\omega,\omega)&=\frac{1}{6 L^2 \omega^2}\Big(3 L^2 (-f (f+1) M+2 f N+N)\nonumber\\
  &-4 \pi^2
   l_B^2 \big(2 f (f+1) (2 f+1) M^3+6 (2 f+1)
    M N^2\nonumber\\
  &-3 N (2 f   M+M)^2-4 N^3+N\big)\Big)
\end{align}
The first line is independent of $L$ and the
geometry of the sample.  The remaining terms both depend on $B$ and $L$. 
Since $\omega$ and $B$ are linearly related,
error propagation leads to the same relative minimal standard deviation of an
unbiased estimator $\sigma(\hat{B})$ of $B$ as for $\omega$. 
Together with Eq.~\eqref{eq:qcrb} we obtain,
\begin{equation}
  \label{eq:dBn0}
  \sigma(\hat{B})\ge 
  \frac{m_\text{eff}/e}{\sqrt{M_e I(\ket{\bm n}_\omega,\omega)}}\,,
\end{equation}
where $M_e$ is the number of independent measurements. A necessary condition for the application of the formula is 
that the very description of the system in terms of harmonic oscillators is adequate.  This implies that the magnetic field must not be too weak, i.e.~$l_B\ll w$, 
which sets a lower bound on $B$ for given 
$w$, $B\gg \hbar/(e w^2)$ with numerical values $B[T]\gg 6.58\times
10^{-16}/(w[\text{m}])^2$. Conversely, for given $B$ the formula
implies a minimal sensor size of  
$w>25.7\,n$m/($\sqrt{B\text{[T]}}$). Secondly, we recall that we assumed $w\gg L$.
In the opposite case, $w$ and $L$ should be exchanged. As explained
above, symmetry under
exchange of $w$ and $L$ is not to be expected in a sensor where the
confining potential breaks that symmetry.\\

The most interesting regime  corresponds to $N\gg 1$.  In a realistic
sample, the areal density 
$n_\text{2D}=N/(L\,w)$ of the electrons is fixed.  In this case, for increasing $N$, $w$ or $L$ must
increase as well, and with them $M\propto L\,w\propto N$. Hence, in the
limit of large $N$ one should replace $M$ by its value
$(N-\mbar)/f$. Suppose $w=\mu N^\lambda l_B$ and $L=\nu N^{1-\lambda}
l_B$ with $1/2\ge \lambda\leq 1$ to
ensure that $w\gg L$ for large $N$. For $\lambda>1/2$, one always has
$w\gg L$ for $N\to\infty$, whereas for $\lambda= 1/2$, $w/L=\mu/\nu$ is
fixed but can be large.   Then the leading term of the QFI becomes  
\begin{equation}
 I(\ket{\bm n}_\omega,\omega)\simeq \frac{2\pi^2}{3f^2\omega^2\nu^2}N^{1+2\lambda}.
\end{equation}
and signals faster 
than ``Heisenberg scaling'' of the QFI (meaning $I(\ket{\bm
  n}_\omega,\omega)\propto N^2$) for $\lambda>1/2$.
The fastest possible scaling, $I(\ket{\bm
  n}_\omega,\omega)\propto N^3$ can 
be achieved in the limit of
fixed $L$ and correspondingly $w\propto N$, i.e.~in the limit of a
strip-like sensor.  

The origin of this super-Heisenberg scaling can be traced back to the
$\sum_m m^2$ in \eqref{eq:Ifinal4} with bounds $\sim M\sim N$ which
gives a scaling $\propto N^3$.  It
arises from occupying high-momentum states in $x$-direction, as is
required for Fermions by the Pauli principle.  Since $k_m$ determines
also the shift $y_m=k_m l_B^2$ of the harmonic oscillators in $y$
direction, large values of $k_m$ lead to correspondingly large
sensitivity to a change of $l_B$ and hence of $B$. Interestingly, if
the kinetic energy in $x$-direction had a power-law scaling with
$k_x$ with a different power, also the scaling of the sensitivity with
$N$ would change, with higher powers being favorable. % Conversely, for graphene-based sensors
% with kinetic energy $\propto |\bm k|$, a power-law of the QFI with the
% power reduced by 1 is expected.  %\qq
Numerically, using typical parameter values for Gallium arsenide
($m_\text{eff}\simeq 0.068m_\text{e}$, $n_\text{2D}\simeq 1.0\times
10^{15}/\text{m}^2)$  and a magnetic field of 1\,T, the
minimal predicted error is of the order of $6.2\times 10^{-11}$\,T for
a single measurement 
with a sensor of size $w=1$\,cm, $L=1$\,mm. The number of
information-carrying  measurements per second is determined by the
bandwidth of the interrogation scheme.  That bandwidth is ultimately
limited by the cyclotron frequency, and hence the number of
measurements in 1\,s cannot be greater
than $M_e \sim 1/(\omega \cdot 1s)\simeq 10^{12}$ at 1\,T.  A more
conservative bandwidth of 10 G\,Hz yields a bound on the achievable
sensitivity of the order of $10^{-16}$\,T/Hz$^{1/2}$, to be compared with 100\,nT/Hz$^{1/2}$ sensitivity of a silicon-based commercially available sensor \cite{Roelver2013}, or another one with $0.4\,\mu$T sensitivity at fields up to about mT with a chip of linear size $\sim$mm \cite{BoschSensor}.  \\

\begin{figure}
\includegraphics[scale=0.5]{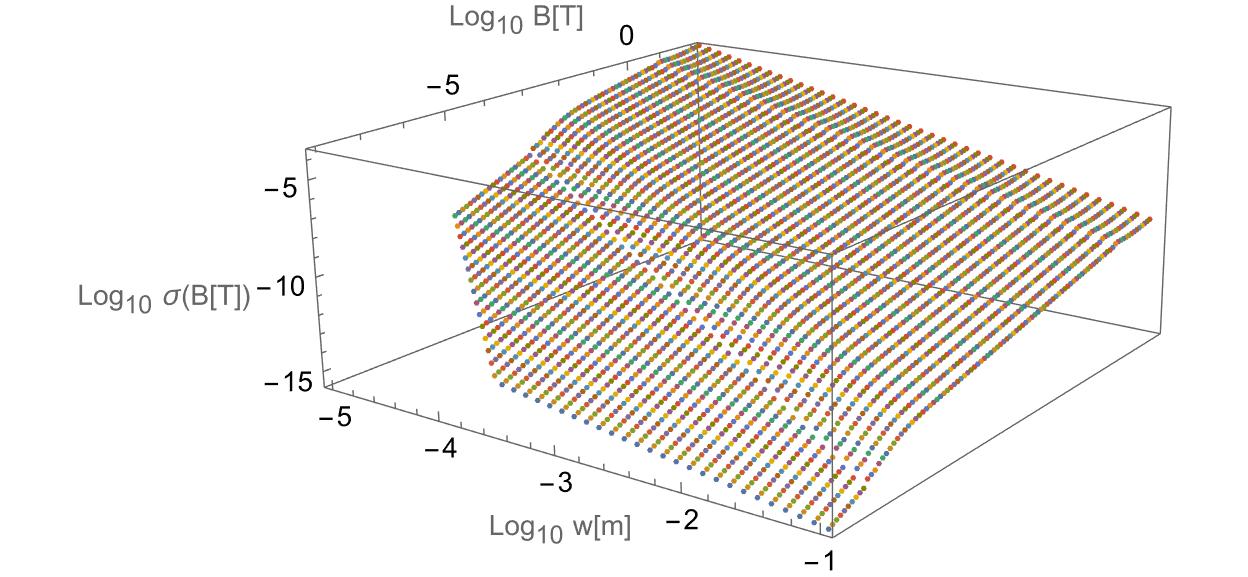}
\caption{Minimal standard deviation of an unbiased estimate of the
  magnetic field $B$ based on the quantum Hall effect,
  Eqs.\eqref{eq:dBn0} and \eqref{eq:Ifinal5} for a single readout, $M_e=1$, as function of the   width $w$ of the sensor and magnetic field $B$ for $10^{-8}\le
  B[T]\le 10^2$. The minimal error is
  plotted only where the theory is applicable, $w>L,l_B(B)$.  
  The length is set to $L=1\mu$\,m, and $w\ge 10^{-5}$\,m.}
  \label{fig1}
\end{figure}

In Fig.~\ref{fig1} we plot the minimal estimation error $\sigma(B)$ as
function of $B$ and 
$w$ in the parameter ranges where 
$l_B(B)<w$ is satisfied.
It should be kept in mind that {\em i.)} we considered the case of
zero temperature and neglected decoherence, and {\em ii.)} the QCRB
provides an idealized lower bound on the error that can rarely be
achieved in practice due to additional technical noise and other
imperfections.  Nevertheless, the QCRB \eqref{eq:dBn0} constitutes an
important benchmark that allows one to understand what sensitivity is
possible in principle as function of $B$ and the size of the sensor.

%*************************************************************************
%*************************************************************************
\section{Conclusion}\label{conclusion}
%*************************************************************************
%*************************************************************************
In summary, we have derived analytical expressions for the quantum
Fisher information (QFI) that determines the smallest possible fluctuations
of an unbiased estimator of a parameter encoded in an arbitrary pure quantum state
of a fermionic many-body system via three different types of unitary
transformations. In the case of two concatenated unitaries we obtained
a simple chain rule for the QFI, Eq.~\eqref{eq:QFIchain}, that
simplifies further for parameters coded through a Bogoliubov
transformation \eqref{eq:genH2bp}, for a many-body basis state \eqref{Ibasisstate}, or a
Hamiltonian time
evolution paired with a modification of the single-particle energy
eigenstates \eqref{eq:QFIUTfin2}.  In the latter case, a variance of a
Hermitian generator naturally arises just as for a single unitary
evolution, albeit taken in an intermediate state.  We applied the general results to
the quantum Hall effect in the ground state of non-interacting
electrons and calculated the smallest possible standard deviation
of an unbiased estimator of the magnetic field. 
We found a scaling of the sensitivity (standard deviation) with which the magnetic field can be measured as 
$1/\sqrt{N^{1+2\lambda}}$, where $\lambda\in[1/2,1]$ 
controls the scaling of the width and length of the sensor with the
number of electrons.

For any $\lambda>1/2$ this corresponds to a super-Heisenberg scaling
of the sensitivity.  It has its physical origin in the occupation of
high-energy momentum states, as required by the Pauli principle, which
lead to large spatial  displacements of the energy eigenstates corresponding to
the Landau-levels, proportional to the momentum and the magnetic
length squared.  The large momenta hence translate to high sensitivity to changes of the
magnetic length and as a consequence of the magnetic field.
It should be kept in mind, however, that the analysis is highly
idealized: zero temperature was assumed, and all decoherence effects
as well as technical noise are neglected.  Future work will have to
show how robust the large sensitivities are, and how they change when
using different materials. 

%************************************************************************************************************************
\section*{Acknowledgments.}
%************************************************************************************************************************

DB thanks OG and the University Paris-Saclay for hospitality for a
stay during which part of this work was done.

\appendix

%*************************************************************************
%*************************************************************************
\section{Alternative proof of \protect(\ref{eq:genH}\protect)}\label{app1}
%*************************************************************************
%*************************************************************************
Here we sketch an alternative proof of \eqref{eq:genH}. For brevity we
define $\hat{H}=i\hat{S}$, and a $2M\times 2M$ matrix $H=iS/2$,
i.e.~$\hat{H}=\alpha H \alpha^t$. One then shows in the fermionic case
by direct calculation that
\begin{equation}
  \label{eq:anticomH}
  [\hat{H},\dot{\hat{H}}]=2 \alpha(H,\dot{H})\alpha^t\,,
\end{equation}
where the commutator-like bilinear form $(A,B)$ of two operators is
defined as
\begin{equation}
  \label{eq:AXB}
  (A,B)\equiv A\Xi B-B\Xi A\,
\end{equation}
with $\Xi$ defined in Eq.~\eqref{eq:X}.
Eq.~\eqref{eq:anticomH} generalizes to higher order commutators
$[A,B]_n$ defined recursively through $[A,B]_{n+1}=[A,[A,B]_n]$
and $[A,B]_0=B$, and correspondingly $(A,B)_{n+1}=(A,(A,B)_n)$ and
$(A,B)_0=B$: 
\begin{equation}
  \label{eq:anticomHn}
 [\hat{H},\dot{\hat{H}}]_n=2^n \alpha(H,\dot{H})_n\alpha^t\,.
\end{equation}
Next, one can write the derivative of an exponential of a parameter
dependent operator alternatively as
\begin{equation}
  \label{eq:debHstand}
  \frac{\partial}{\partial \omega}e^{\beta H(\omega)}=\int_0^\beta ds\,
  e^{(\beta-s)H(\omega)}\frac{\partial H(\omega)}{\partial \omega}e^{sH(\omega)}\,,
\end{equation}
where $\beta$ is an arbitrary real number.
The simplest proof of \eqref{eq:debHstand} follows the lines of \cite{wilcox_exponential_1967} by showing
that both sides of the equation satisfy the first-order differential
equation
\begin{equation}
  \label{eq:eqdiff}
  \frac{\partial F}{\partial\beta}-H(\omega) F(\beta)=\frac{\partial
    H(\omega)}{\partial \omega} e^{\beta H(\omega)}\,,
\end{equation}
together with $F(0)=0$, which fixes the solution
uniquely. 
Next one checks the identities
$e^{A \Xi }Be^{-\Xi A}=\sum_{n=0}^\infty\frac{1}{n!}(A,B)_n$  and
$e^ABe^{-A}=\sum_{n=0}^\infty \frac{1}{n!}[A,B]_n$. With this we have
\begin{eqnarray}
  \label{eq:genH3}
  {\mathcal H}&=&-i \hat{T}^\dagger\dot{\hat{T}}=(-i)\int_0^1 du e^{-i u
  \hat{S}}i\frac{\partial\hat{S}}{\partial \omega}e^{-i u
  \hat{S}}\\
&=&\int_0^1du\,\sum_{n=0}^\infty\frac{1}{n!}[-iu\hat{S},\dot{\hat{S}}]_n\\
&=&\frac{\alpha}{2}\left[
\int_0^1
    du\,\sum_{n=0}^\infty\frac{(-iu)^n}{n!}(S,\dot{S})_n\right]\alpha^t\\
&=&\frac{\alpha}{2}\int_0^1du\,e^{-iu
    S\Xi}\dot{S} e^{iu \Xi S}\alpha^t\\
&=&\frac{1}{2}\alpha\tilde{\Omega}\alpha^t
\end{eqnarray}
with $\tilde{\Omega}$ given by \eqref{eq:Omegat}, which completes the
proof. \\

%*************************************************************************
%*************************************************************************
\section{Derivation of wavefunction overlaps}\label{calculfinaloverlap}
%*************************************************************************
%*************************************************************************
We start from Eq.~\eqref{eq:overlapom}.
Instead of varying the frequency, it is more convenient to make the change of variables to the magnetic length $l\equiv l_B=\sqrt{\hbar/(m_{\rm eff}\omega)}$ (for ease of notation, in the present Appendix we denote the magnetic length just by $l$). Equation \eqref{eq:overlapom} becomes
\begin{equation}\label{eq:overlap}
 _{\omega_0}\langle n',m'|n,m\rangle_{\omega_0+\delta\omega}\simeq \delta_{n,n'}\delta_{m,m'} +\ _{l}\langle n',m'|\partial_{l}|n,m\rangle_{l}\frac{dl}{d\omega_0}\delta\omega\,,
\end{equation}
and $dl/d\omega_0=-(1/2)l/\omega_0$. The matrix element
\begin{equation}\label{eq:ME}
  _{l}\langle n',m'|\partial_{l}|n,m\rangle_{l}=\int dx dy\,\, {}_l\langle n',m'|x,y\rangle \partial_l \langle x,y|n,m\rangle_l 
\end{equation}
can now be calculated from the explicit expression \eqref{eq:WF} of the wave functions. In particular \eqref{eq:WF} gives
\begin{equation}
 \partial_l \langle x,y|n,m\rangle_l  = \frac{e^{i k_m x}}{\sqrt{L}}\partial_l \chi_{n,m}(\eta)
\end{equation}
with
\begin{equation}\label{eq:eta}
 \eta=\frac{y}{l}-k_ml.
\end{equation}
Integration over $x$ yields a $\delta_{m,m'}$ coefficient.
The matrix element \eqref{eq:ME} becomes
\begin{equation}\label{eq:ME2}
  _{l}\langle n',m'|\partial_{l}|n,m\rangle_{l}=\delta_{m,m'}l \int d\eta\, \chi_{n',m'}(\eta)\partial_l\chi_{n,m}(\eta)\,.
\end{equation}
Using
\begin{equation}
 \frac{d\eta}{dl} = -\frac{y}{l^2} -k_m = -\frac{\eta}{l} - 2k_m,
\end{equation}
we get
\begin{widetext}
\begin{equation}
 \partial_l \chi_{n,m}(\eta) = -\frac{\chi_{n,m}(\eta)}{2l} + \frac{{\cal N}_n}{\sqrt{l}}\frac{d\eta}{dl}\partial_{\eta}\left[H_n(\eta)e^{-\eta^2/2}\right]
 = -\frac{\chi_{n,m}(\eta)}{2l} - \frac{{\cal N}_n}{\sqrt{l}} \left(\frac{\eta}{l}+2k_m\right)\left[\partial_{\eta}H_n(\eta)-\eta H_n(\eta)\right]e^{-\eta^2/2}\,.
\end{equation}
One thus obtains for the matrix element (\ref{eq:ME})
\begin{align}
  _{l}\langle n',m'|\partial_{l}|n,m\rangle_{l} &= - \frac{1}{2l}\delta_{n,n'}\delta_{m,m'}- \delta_{m,m'}{\cal N}_n {\cal N}_{n'}\int d\eta\, e^{-\eta^2}H_{n'}(\eta)   \left(\frac{\eta}{l}+2k_m\right) \left[\partial_{\eta}H_n(\eta)-\eta H_n(\eta)\right]\nonumber\\
  &= - \frac{1}{2l}\delta_{n,n'}\delta_{m,m'}- \delta_{m,m'}{\cal N}_n {\cal N}_{n'}\int d\eta\, e^{-\eta^2}H_{n'}(\eta) 
  \left(\frac{\eta}{l}+2k_m\right) \left[2nH_{n-1}(\eta)-\eta H_n(\eta)\right]\nonumber\\
  &= - \frac{1}{2l}\delta_{n,n'}\delta_{m,m'}\nonumber\\
  & - \delta_{m,m'}{\cal N}_n {\cal N}_{n'}\int d\eta\, e^{-\eta^2}H_{n'}(\eta) \left[4k_m n H_{n-1}(\eta) 
  + \frac{2n}{l}\eta H_{n-1}(\eta) -  2k_m\eta H_n(\eta) - \frac{\eta^2}{l} H_n(\eta)  \right],
\label{ldll}
\end{align}
where we have used $\partial_{\eta}H_n(\eta)= 2n H_{n-1}(\eta)$ \cite{abramowitz1964handbook}. These integrals can be evaluated if we express $\eta$ and $\eta^2$ in terms of Hermite
polynomials,
\begin{equation}
 H_1(\eta)=2\eta, \qquad H_2(\eta)=2(2\eta^2 - 1)\Leftrightarrow \eta^2=\frac{1}{2}+\frac{1}{4}H_2(\eta),
\end{equation}
so that \eqref{ldll} becomes
\begin{eqnarray}\label{eq:ME3}
%\nonumber
  _{l}\langle n',m'|\partial_{l}|n,m\rangle_{l} &=&  
  - \frac{1}{2l}\delta_{n,n'}\delta_{m,m'} - \delta_{m,m'}{\cal N}_n {\cal N}_{n'}\int d\eta\, e^{-\eta^2}H_{n'}(\eta)\left[4k_m n H_{n-1}(\eta) \right.\\
  \nonumber
  && + \frac{n}{l}H_1(\eta) H_{n-1}(\eta) - \left. k_m H_1(\eta) H_n(\eta) - \frac{1}{2l} H_n(\eta) -\frac{1}{4l}H_2(\eta)H_n(\eta) \right]\,.
\end{eqnarray}
We can now use the identity \cite{abramowitz1964handbook}
\begin{equation}\label{eq:herm}
 \int_{-\infty}^{\infty} d\eta\, e^{-\eta^2}H_{n'}(\eta)H_m(\eta)H_n(\eta) = \frac{2^{(m+n+n')/2}\sqrt{\pi}n'!n!m!}{(s-n')!(s-n)!(s-m)!},
\end{equation}
where $s=\frac12(n+n'+m)$ (note that $s$ is an integer due to the parity of the Hermite polynomials -- indeed, for an odd integer value of $n+n'+m$ the integrand in \eqref{eq:herm} is an odd function and thus the integral vanishes). We also make use of the 
orthogonality relation of the Hermite polynomials
\begin{equation}
 \int_{-\infty}^{\infty} d\eta\, e^{-\eta^2}H_{n'}(\eta)H_n(\eta)= \sqrt{\pi}2^n n! \delta_{n',n},
\end{equation}
so that the matrix element \eqref{eq:ME3} becomes
\begin{align}
 \nonumber
& _{l}\langle n',m'|\partial_{l}|n,m\rangle_{l} =
  - 2\sqrt{2n}k_m \delta_{n',n-1}\delta_{m',m}- \delta_{m',m} {\cal N}_n {\cal N}_{n'} \left[ \frac{n}{l} \frac{2^{(n+n')/2}\sqrt{\pi}n'!(n-1)!}
  {\left(\frac{n'-n}{2}+1\right)!\left(\frac{n-n'}{2}\right)!\left(\frac{n+n'}{2}-1\right)!}\Pi(n+n')\right.  \nonumber\\
  &\left.- k_m\frac{2^{(n+n'+1)/2}\sqrt{\pi}n'!n!}{\left(\frac{n+1-n'}{2}\right)!\left(\frac{n'+1-n}{2}\right)!\left(\frac{n+n'-1}{2}\right)!}\Pi(n+n'+1)
  -\frac{1}{4l}\frac{2^{2+(n+n')/2}\sqrt{\pi}n'!n!}{\left(\frac{n-n'}{2}+1\right)!\left(\frac{n'-n}{2}+1\right)!\left(\frac{n+n'}{2}-1\right)!}
  \right]\Pi(n+n')\,,
\label{Pipi}
\end{align}
where we have introduced a parity function 
\begin{equation}
 \Pi(m)=\left\{
 \begin{array}{ccc}
  1 & {\rm for} & m~~{\rm even}\\
  0 & {\rm for} & m~~{\rm odd}
 \end{array}\right.
\end{equation}
to take into account the different cases for which the integral (\ref{eq:herm}) vanishes. Further simplifications can be obtained by noticing that the factorials in the denominators of \eqref{Pipi} involve both $n'-n$ and $n-n'$ in each term. Since the factorial of a negative number is infinite, we get the following simplifications
\begin{eqnarray}
\frac{\Pi(n+n')}{\left(\frac{n'-n}{2}+1\right)!\left(\frac{n-n'}{2}\right)!}&=&\delta_{n',n-2}+\delta_{n',n}\\
\frac{\Pi(n+n'+1)}{\left(\frac{n+1-n'}{2}\right)!\left(\frac{n'+1-n}{2}\right)!}&=&\delta_{n',n-1}+\delta_{n',n+1}\\
\frac{\Pi(n+n')}{\left(\frac{n-n'}{2}+1\right)!\left(\frac{n'-n}{2}+1\right)!}&=&\frac12\delta_{n',n-2}+\delta_{n',n}+\frac12\delta_{n',n+2}\,.
\end{eqnarray}
Replacing the normalization factors ${\cal N}_n{\cal N}_{n'}$ by their value \eqref{Nn}, Eq.~\eqref{Pipi} then reduces to
\begin{eqnarray}
 \nonumber
 _{l}\langle n',m'|\partial_{l}|n,m\rangle_{l} &=&  
 % \frac{1}{l}\delta_{n',n}\delta_{m',m} 
  - 2\sqrt{2n}k_m \delta_{n',n-1}\delta_{m',m} \\ 
  \nonumber
  &&- \delta_{m',m}\sqrt{n'!n!} \left[ \frac{1}{2l} \frac{1}{\left(\frac{n+n'}{2}-1\right)!}\left(\delta_{n',n-2}-\delta_{n',n+2} \right)
  \right.\\
  &&
  -\left.  \sqrt{2}k_m\frac{1}{\left(\frac{n+n'-1}{2}\right)!} \left(\delta_{n',n-1}+\delta_{n',n+1}\right) \right]\,,
\end{eqnarray}
which further simplifies to
\begin{eqnarray}
 \nonumber
 _{l}\langle n',m'|\partial_{l}|n,m\rangle_{l} &=&  
 -\sqrt{2} k_m\left(\sqrt{n}\delta_{n',n-1}-\sqrt{n+1}\delta_{n',n+1}\right)\delta_{m',m}\nonumber\\
  &-&\frac{1}{2l} \left(\sqrt{n(n-1)}\delta_{n',n-2}-\sqrt{(n+2)(n+1)}\delta_{n',n+2} \right)\delta_{m',m}\,. \end{eqnarray}
Expanding expression (\ref{eq:overlap}) to linear order in $\delta\omega$ we directly get Eq.~\eqref{finaloverlap}.\\

%*************************************************************************
%*************************************************************************
\section{Alternative derivation of \protect(\ref{finaloverlap}\protect)}
%*************************************************************************
%*************************************************************************
\label{otherproofoverlap}
We start from the Hutchisson formula for the overlaps
$R_{nn'}={_{\omega_0}}\braket{n}{n'}_\omega$. It is given by
\cite{smith_overlap_1969} with $\nu'=\omega_0$ and
$\nu''=\omega$, and reads
\begin{equation}
\label{defR}
R_{nn'}=\sqrt{2^{-(n+n')}qn!n'!}(-1)^{n'}e^{-\frac14\gamma^2p}
\sum_{r=0}^{\min(n,n')}
\frac{(-2q)^r}{r!}\sum_{s=0}^{\lfloor (n'-r)/2\rfloor}\frac{(\gamma p)^{n'-r-2s}}{(n'-r-2s)!}\frac{x^s}{s!}\sum_{t=0}^{\lfloor (n-r)/2\rfloor}\frac{(\gamma q)^{n-r-2t}}{(n-r-2t)!}\frac{(-x)^t}{t!},
\end{equation}
with $\gamma=\sqrt{\omega m_{\rm eff}/\hbar}d=d/l_B(\omega)$,
$d=k_m(l_B^2(\omega)-l_B^2(\omega_0))$,
$x=(\omega-\omega_0)/(\omega+\omega_0)$,
$q=2(\omega\omega_0)^{1/2}/(\omega+\omega_0)$, and
$p=2\omega_0/(\omega+\omega_0)$. 

We want to calculate the first derivative of $R_{nn'}$ with respect to $\omega$, taken at $\omega=\omega_0$. In that limit we have  $(\gamma^2 p)'=q'=0$. Contributions to the derivative will therefore come from derivatives of terms of the form $(\gamma p)^a(\gamma q)^bx^c$ with $a,b,c\geq 0$, that is,
\begin{equation}
\label{3terms}
[(\gamma p)^a(\gamma q)^bx^c]'=a(\gamma p)^{a-1}(\gamma p)'(\gamma q)^bx^c
+(\gamma p)^ab(\gamma q)^{b-1}(\gamma q)'x^c+(\gamma p)^a(\gamma q)^bcx^{c-1}x'.
\end{equation}
\end{widetext}
In the limit  $\omega=\omega_0$ we have $\gamma=x=0$, so that only terms with an exponent 0 in \eqref{3terms} can yield a nonzero contribution. Equation \eqref{3terms} reduces to 
$(\gamma p)'=-k_m/\omega_0^{3/2}$ for $a=1, b=c=0$, to $(\gamma q)'=-k_m/\omega_0^{3/2}$ for $b=1,a=c=0$, and to $x'=1/(2\omega_0)$ for $c=1,a=b=0$.
The first case corresponds to $s=t=0$, $r=n'-1=n$. The second case gives $s=t=0$ and $r=n'=n-1$. The third case leads to either $s=0,t=1,r=n'=n-1$ or $s=1,t=0,r=n'-1=n$. Gathering all contributions together we exactly get the first-order term of Eq.~\eqref{finaloverlap}.

\bibliography{mybibs_bt_170418}

\end{document}